 \documentclass[10pt,preprint]{aastex}  %compact preprint style (MJA)
 \usepackage{natbib}
\def\ang{\AA}

\def\gapprox{\lower.4ex\hbox{$\;\buildrel >\over{\scriptstyle\sim}\;$}}
\def\lapprox{\lower.4ex\hbox{$\;\buildrel <\over{\scriptstyle\sim}\;$}}

\shortauthors{ASCHWANDEN ET AL. 2015}
\shorttitle{Global Energetics of Solar Flares. V.}

\begin{document}
%{\sl  Manuscript, accepted ... }

\title{         Global Energetics of Solar Flares and CMEs: 
		V. Energy Closure }

\author{        Markus J. Aschwanden$^1$, 
	        Amir Caspi$^2$,
	        Christina M.S. Cohen$^3$,
	        Gordon Holman$^4$,
	        Ju Jing$^5$,
	        Matthieu, Kretzschmar$^6$,
	        Eduard P. Kontar$^7$,
	        James M. McTiernan$^8$,
		Richard A. Mewaldt$^3$,
	        Aidan O'Flannagain$^9$,
	        Ian G. Richardson$^{10}$,
 	        Daniel Ryan$^{11}$,
	        Harry P. Warren$^{12}$,
	        Yan Xu$^{13}$}

\affil{		$^1)$ Lockheed Martin, 
		Solar and Astrophysics Laboratory, 
                Org. A021S, Bldg.~252, 3251 Hanover St.,
                Palo Alto, CA 94304, USA;
                e-mail: aschwanden@lmsal.com }

\affil{         $^2)$ Planetary Science Directorate,
                Southwest Research Institute,
                Boulder, CO 80302, USA;
                e-mail: amir.caspi@swri.org }

\affil{		$^3)$ California Institute of Technology,
		Mail Code 290-17,
		Pasadena, CA 91125, USA;
		e-mail: rmewaldt@srl.caltech.edu.}

\affil{         $^4)$ Code 671, NASA Goddard Space Flight Center,
                Greenbelt, MD 20771, USA;
                e-mail: gordon.d.holman@nasa.gov }

\affil{         $^5)$ Space Weather Research Laboratory,
                Center for Solar-Terrestrial Research,
                New Jersey Institute of Technology,
                323 Martin Luther King Blvd., Newark, NJ 07102-1982, USA;
                e-mails: ju.jing@njit.edu }

\affil{		$^6)$ LPC2E, UMR 6115 CNRS and University of Orl\'eans,
		3a Av. de la recherche scientifique,
		45071 Orl\'eans, France;
		e-mal: matthieu.kretzschmar@cnrs-orleans.fr}

\affil{         $^7)$ School of Physics and Astronomy, University of Glasgow,
                G12 8QQ, Glasgow, Scotland, UK;
                e-mail: eduard.kontar@astro.gla.ac.uk }

\affil{         $^8)$ Space Sciences Laboratory,
                University of California,
                Berkeley, CA 94720, USA;
                e-mail: jimm@ssl.berkeley.edu }

\affil{         $^9)$ Astrophysics Research Group,
                School of Physics, Trinity College Dublin,
                Dublin 2, Ireland;
                e-mail: aidanoflann@gmail.com}

\affil{         $^{10})$ CESST and Dept. Astronomy,
		University of Maryland;
		Code 661, NASA Goddard Space Flight Center
		Greenbelt, MD 20770, USA;
                e-mail: richardson@lheavx.gsfc.nasa.gov}

\affil{         $^{11})$ NASA Goddard Space Flight Center,
		8800 Greenbelt Road, 
		Greenbelt, MD 20770, USA;
                e-mail: ryand5@tcd.ie }

\affil{         $^{12})$ Space Science Division,
                Naval Research Laboratory,
                Washington, DC 20375, USA;
                e-mail: harry.warren@nrl.navy.mil }

\affil{         $^{13})$ Space Weather Research Laboratory,
                Center for Solar-Terrestrial Research,
                New Jersey Institute of Technology,
                323 Martin Luther King Blvd., Newark, NJ 07102-1982, USA;
                e-mails: yan.xu@njit.edu }

\begin{abstract}
In this study we synthesize the results of four previous studies
on the global energetics of solar flares and associated coronal 
mass ejections (CMEs), which include magnetic, thermal, nonthermal, 
and CME energies in 399 solar M and X-class flare events observed
during the first 3.5 years of the Solar Dynamics Observatory (SDO) 
mission. Our findings are: (1) The sum of the mean nonthermal energy 
of flare-accelerated particles ($E_{\mathrm{nt}}$), the energy of direct heating 
($E_{\mathrm{dir}}$), and the energy in coronal mass ejections ($E_{\mathrm{CME}}$), 
which are the primary energy dissipation processes in a flare, is found 
to have a ratio of $(E_{\mathrm{nt}}+E_{\mathrm{dir}}+
E_{\mathrm{CME}})/E_{\mathrm{mag}} = 0.87 \pm 0.18$, 
compared with the dissipated magnetic free energy $E_{\mathrm{mag}}$, which confirms 
energy closure within the measurement uncertainties and corroborates the
magnetic origin of flares and CMEs;
(2) The energy partition of the dissipated magnetic free energy is: 
$0.51\pm0.17$ in nonthermal energy of $\ge 6$ keV electrons, 
$0.17\pm0.17$ in nonthermal $\ge 1$ MeV ions, $0.07\pm0.14$ in
CMEs, and $0.07\pm0.17$ in direct heating; 
(3) The thermal energy is almost always less 
than the nonthermal energy, which is consistent with the thick-target model; 
(4) The bolometric luminosity in white-light flares is comparable with 
the thermal energy in soft X-rays (SXR); (5) Solar Energetic Particle (SEP) 
events carry a fraction $\approx 0.03$ of the CME energy, which is consistent 
with CME-driven shock acceleration; 
and (6) The warm-target model predicts a lower limit of the
low-energy cutoff at $e_c \approx 6$ keV, based on the mean differential 
emission measure (DEM) peak temperature of $T_e=8.6$ MK during flares. 
This work represents the first statistical study that establishes energy 
closure in solar flare/CME events.
\end{abstract}

\section{	INTRODUCTION			}

Energy closure is studied in many dynamical processes, such as in
meteorology and atmospheric physics (e.g., the turbulent kinetic TKE 
and potential energies TPE make up the turbulent total energy, 
TTE = TKE + TPE; Zilitinkevich et al.~2007), in magnetospheric and
ionospheric physics (e.g., where the solar wind transfers energy
into the magnetosphere in form of electric currents;
Atkinson, 1978), or in astrophysics (e.g., in the
energetics of {\sl Swift} gamma-ray burst X-ray afterglows;
Racusin et al.~2009). The most famous example is probably the
missing mass needed to close our universe (e.g., White et al.~1993).
Here we investigate the energy closure
in solar flare and coronal mass ejection (CME) events, which
entail dissipated magnetic energies (Aschwanden, Xu, and Jing 2014; 
Paper I), thermal energies (Aschwanden et al.~2015a; Paper II), 
nonthermal energies (Aschwanden et al.~2016; Paper III), and
kinetic and gravitational energies of CMEs (Aschwanden 2016a; Paper IV). 

The energy flow in solar flares and CMEs passes through several 
processes which are depicted in the diagram of Fig.~1. Initially,
a stable non-flaring active region exists with a near-potential
magnetic field with energy $E_p$, which then becomes twisted and 
sheared, building up nonpotential energy $E_{\mathrm{np}}$ and the free 
energy, $E_{\mathrm{free}}=E_{\mathrm{np}}-E_{p}$, of which a fraction 
$E_{\mathrm{mag}} \le E_{\mathrm{free}}$ 
is dissipated during a flare (e.g., Schrijver 
et al.~2008; Aschwanden 2013). There are three primary energy
dissipation processes that follow after a magnetic instability,
typically a magnetic reconnection process, spawning (1) the acceleration
of nonthermal particles (e.g., reviews by Miller et al.~1997;
Aschwanden 2002; Benz 2008; Holman et al.~2011), 
with electron energy $E_{\mathrm{nt,e}}$ and
ion energy $E_{\mathrm{nt,i}}$, providing (2) direct heating in
the magnetic reconnection region, $E_{\mathrm{dir}}$ 
(e.g., Sui and Holman 2003; Caspi and Lin 2010; Caspi et al.~2015), 
and are often accompanied by (3) an eruptive process, 
which can be a complete eruption of a CME or filament, or a semi-eruptive
energy release, also known as ``failed eruption'', in the case of a 
confined flare (e.g., T\"or\"ok and Kliem 2005). The CME process carries
an energy of $E_{\mathrm{CME}}=E_{\mathrm{kin}} + E_{\mathrm{grav}}$, consisting
of the kinetic energy $E_{\mathrm{kin}}$ and the gravitational potential energy
$E_{\mathrm{grav}}$, to lift a CME from the solar surface into the
heliosphere. These primary energy dissipation processes allow us
to test the primary energy closure equation,
\begin{equation}
	E_{\mathrm{mag}} = 
	( E_{\mathrm{nt}} + E_{\mathrm{dir}} + E_{\mathrm{CME}} ) = 
	( E_{\mathrm{nt,e}} + E_{\mathrm{nt,i}} + E_{\mathrm{dir}} + 
          E_{\mathrm{CME,kin}} + E_{\mathrm{CME,grav}} ) \ ,
\end{equation}
where the left-most side of the equation contains the total (magnetic) 
energy input (or storage), and the right-most side of the equation 
contains the total energy output (or dissipation).

After this primary step in the initiation of a flare and CME,
secondary energy dissipation processes kick in. Nonthermal particles
are accelerated along bi-directional trajectories that lead out of 
the magnetic reconnection region, where most particles precipitate
down to the chromosphere, heat chromospheric plasma and
drive evaporation of the heated plasma up into the corona
(e.g., Antonucci and Dennis 1983),
while other particles escape into interplanetary
space (see reviews by Hudson and Ryan 1995; Aschwanden 2002; Lin 2007). 
The flare arcade that becomes filled with heated 
chromospheric plasma radiates and loses its energy by conduction 
and radiation in soft X-rays (SXR) and extreme ultra violet (EUV). 
The thermal energy
content $E_{\mathrm{th}}$ can be calculated from the total emission measure
observed in SXR and EUV and should not exceed the nonthermal
energy, $E_{\mathrm{nt}} = E_{\mathrm{nt,e}} + E_{\mathrm{nt,i}}$,
unless there are other heating processes besides the electron beam-driven
heating observed in hard X-rays (according to the thick-target 
bremsstrahlung model of Brown 1971). Thus we can test the following energy 
inequality between thermal and nonthermal energies (if we neglect direct
heating),
\begin{equation}
	E_{\mathrm{th}} \le E_{\mathrm{nt}} 
	= ( E_{\mathrm{nt,e}} + E_{\mathrm{nt,i}} ) \ .
\end{equation}
Radiation is not only produced at SXR and EUV wavelengths ($E_{\mathrm{th}}$), 
but also in visible and near-ultraviolet wavelengths, recorded 
as white-light flare emission, being the largest contributor to the 
bolometric energy or luminosity 
$E_{\mathrm{bol}}$, which contains vastly more radiative energy 
than observed in SXR (Woods et al.~2004, 2006; Kretzschmar 2011). 
Using a superimposed epoch analysis of 2100 C-, M-, and X- class flares, 
Kretzschmar (2010; 2011 and Table 1 therein) calculated the total solar 
irradiance (TSI) for five synthesized flare time profiles. 
The so determined
continuum emission produced by white-light flares allows us 
to compare another pair of energies, which relates the total
thermal energy $E_{\mathrm{th}}$ to the bolometric luminosity, 
produced by the flare impact of precipitating particles, 
radiative back-warming, and locally enhanced ionization, enhancing
bound-free and free continuum emission (e.g., 
Najita and Orrall 1970; Hudson 1972; Ding et al.~2003;
Battaglia and Kontar 2011; Battaglia et al.~2011; Xu et al.~2014),
\begin{equation}
	E_{\mathrm{bol}} \approx E_{\mathrm{th}} \ .
\end{equation}

Another secondary process is the acceleration of nonthermal particles
by the CME, which is produced by shock acceleration in very fast 
CMEs, observed in form of solar energetic particle (SEP)
events (e.g., see review by Reames 2013), 
which allows us to test another energy inequality,
\begin{equation}
	E_{\mathrm{SEP}} \le E_{\mathrm{CME}} = 
	E_{\mathrm{CME,kin}} + E_{\mathrm{CME,grav}} \ .
\end{equation}

The energy closure studied here depends, of course, on specific
physical models of flares and CMEs. Here we discuss only the most
common solar flare models, but we have to make a disclaimer that
alternative flare models may deviate from the energy closure
relationships and inequalities discussed here. Another important
issue in any energy closure relationship concerns the double-counting
of energies if there are multiple energy conversion processes acting
at the same time or near-simultaneously. We attempt to distinguish
between primary and secondary energy dissipation mechanisms,
as shown in Fig.~1. 

The aim of this paper is to summarize the assumptions that went
into the derivation of the various measured and observationally
derived energy parameters (Section 2), to test energy closure
(Section 3), to discuss some physical processes that play a role 
in the energy closure relationships (Section 4), and final 
conclusions (5).

\section{		FLARE AND CME ENERGIES 			}

In order to characterize the different forms of energies that can
be measured or derived in solar flares and CMEs we start with a
brief description of the basic assumptions that are made in the
four relevant studies (Papers I, II, III, IV, and references
therein) in the derivation of various forms of energies. 

We will quote the mean ratios of the various energy conversion
processes $E_x$ to the dissipated magnetic energy $E_{\mathrm{mag}}$,
by averaging their logarithmic values, so that the logarithmic
standard deviation $\sigma_{log}$ corresponds to a factor with respect 
to the mean value. For instance, the ratio of the nonthermal
energy to the magnetically dissipated energy (Section 2.2)
has a logarithmic mean and standard deviation of ${}^{10}log(q_{nt,e})
= {}^{10}log{(E_{nt,e}/E_{\mathrm{mag}})} = -0.39 \pm 0.89$,
which we quote as a linear value with a standard deviation
factor, i.e., 
$q_{nt,e}= (E_{nt,e}/E_{\mathrm{mag}}) = 10^{-0.39} \div 10^{0.89}
= 0.41 \div 7.7$. Thus the range of one standard deviation,
i.e., $[0.41/7.7, 0.41 \times 7.7] = [0.05, 3.2]$, includes
68\% of the events. The statistical error $e_x$ of the mean value 
$q_x$ is then obtained by dividing the standard deviation by the
square root of $n_x$ events. For instance, the error $e_x$ of 
the mean nonthermal energy based on $n=76$ values is 
$e_{nt,e}=10^{0.89/\sqrt{76}} - 1 = 0.26$, given as
$q_x \pm e_x = 0.41\pm0.26$, which expresses the 68\%
statistical probability to find a mean value in this range
for another data set with the same number of $n=76$ events.

\subsection{		Magnetic Energies			}

The basic assumptions in the calculation of magnetic energies are (Paper I):
(1) The coronal magnetic field in a flaring active region is
nonpotential and has a nonpotential energy $E_{\mathrm{np}} = E_p
+ E_{\mathrm{free}}$, with the free energy being larger than zero; 
(2) The free energy $E_{\mathrm{free}}=B_{\varphi}^2/8\pi$ can largely be 
represented by helically twisted fields 
${\bf B} = {\bf B}_r + {\bf B}_\varphi$. It is composed of a
potential field component ${\bf B}_r$ and a nonpotential field component
${\bf B}_{\varphi}$ in perpendicular (azimuthal) direction to the potential field, 
which is induced by vertical currents ${\bf j} (4 \pi/c) = (\nabla \times {\bf B})$
above magnetic field concentrations (such as sunspots or in active region
plages); 
(3) The line-of-sight component $B_z(x,y)$ can be measured from 
magnetograms (such as the {\it Helioseismic and Magnetic Imager (HMI)}; 
(4) The photospheric magnetic field is not force-free and the
transverse magnetic field components cannot be directly measured
from photospheric magnetograms, such as with tradidional
nonlinear force-free field (NLFFF) codes (Aschwanden 2016b);
(5) Coronal loops are embedded in low plasma-$\beta$ regions
and are force-free before and after a flare or CME launch;
(6) The transverse components $B_x(x,y)$ and
$B_y(x,y)$ can be constrained from the 2D directional vectors of coronal
loops observed in EUV wavelengths (such as with the Atmospheric
Imaging Assembly (AIA), using an automated loop tracing algorithm; 
(7) A flare or launch of a CME dissipates a fraction of the
free magnetic energy, and thus the time evolution of the free
energy $E_{\mathrm{free}}(t)$ exhibits in principle a step function 
from a higher (preflare) to a lower (postflare) value of the
free energy;
(8) The evolution of the free energy $E_{\mathrm{free}}(t)$ may 
exhibit an apparent increase due to coronal
illumination effects (such as chromospheric evaporation) at the 
beginning of the impulsive flare phase, before the decrease of 
free energy is observed.

In our global energetics study on magnetic energies (Paper I) 
we included all Geostationary Operational Environmental
Satellite (GOES) X- and M-class flares during the first 3.5
years of the Solar Dynamics Observatory (SDO) mission, 
which amounted to a total data set of
399 flare events. Restricting the magnetic analysis to events
with a longitude difference of $|l-l_0| \le 45^0$ from the 
central meridian ($l_0$) due to foreshortening effects in the
magnetograms, we were able to determine the flare-dissipated
magnetic energy in 172 events, covering a range of 
$E_{\mathrm{mag}} = (1.5 - 1500) \times 10^{30}$ erg.
The dissipated flare energy $E_{\mathrm{mag}}$ is highly correlated with
the free energy $E_{\mathrm{free}}$, the potential energy $E_p$, and the
nonpotential energy $E_{\mathrm{np}}$ (see Fig.~13 in Paper I).

In a previous study on the global flare energetics (Emslie et al.~2012),
no attempt was made to calculate a nonpotential magnetic field
energy change during flares, but instead an {\sl ad hoc} value of 30\% of
the potential energy was assumed. The inferred range of 
$E_{\mathrm{mag}} = (110 - 2900) \times 10^{30}$ erg
appears to over-estimate the dissipated magnetic flare energy by
one to two orders of magnitude for M-class flares, when compared 
with our study. 

\subsection{		Nonthermal Electron Energies		}

The nonthermal energies, which includes the kinetic energy of particles
accelerated out of the thermal population, are derived from hard
X-ray spectra observed with the {\it Ramaty High-Energy Solar Spectroscopic
Imager (RHESSI)} instrument (Paper III). The basic assumptions are:
(1) Particle acceleration occurs in magnetic reconnection processes,
either by electric fields, by stochastic wave-particle interactions
in turbulent plasmas, or by shock waves; 
(2) Hard X-ray spectra are produced by bremsstrahlung (free-free and 
free-bound emission) of both thermal and nonthermal particles; 
(3) The thermal and nonthermal emission can be distinguished in
hard X-ray spectra by an exponential-like spectrum at low energies
(typically 6-20 keV) and a powerlaw-like spectrum at higher energies
(typically 20-50 keV);  
(4) The energy in nonthermal electrons can be calculated by spectral
integration of the powerlaw-like nonthermal spectrum (with a slope
$\delta$) above some low-energy cutoff $e_c$;
(5) The low-energy cutoff $e_c$ can be estimated from the
warm-target model of Kontar et al.~(2015) according to
$e_c = \delta k_B T_e$, where $T_e$ is the average temperature of the
warm-target, in which the electrons diffuse before they lose their energy
by collisions, and $\delta$ is the power law slope of the nonthermal
electron flux;
(6) The warm-target temperature $T_e$ can be estimated from the
mean value of the peak temperature of the differential emission measure
(DEM) distribution observed with {\it AIA} in the temperature range of
$T_e \approx 0.5-20$ MK, which was found to be $T_e=8.6$ MK 
in the statistical average of all events. 
A mean value of the low-energy cutoff $e_c=6.2 \pm 1.6$ keV
was obtained from the entire ensemble of analyzed events (see
discussion in Section 4.1). 

In our global energetics study on nonthermal energies (Paper III)
we analyzed {\it RHESSI} spectra in 191 M- and X-class flare events,
amounting to 48\% of the total data set, while the remainder
was missed due to the duty cycle
of {\it RHESSI} in the day/night portions of the spacecraft orbit.
The nonthermal energies of the 191 analyzed events covers a range of 
$E_{\mathrm{nt,e}} = (0.05 - 8000) \times 10^{30}$ erg.
Cross-correlating the nonthermal energy in electrons $E_{\mathrm{nt,e}}$ with
the dissipated magnetic energy $E_{\mathrm{mag}}$, we find an overlapping
subset of 76 events, which exhibits a mean (logarithmic)
energy ratio of $E_{\mathrm{nt,e}}/E_{\mathrm{mag}} = 0.41 \pm 0.26$, with a standard
deviation factor of $\sigma=7.7$ (Fig.~2a). The distribution
of (logarithmic) electron energies can be represented by a 
log-normal (Gaussian) distribution (Fig.~2b) that extends over a
range of $E_{\mathrm{mag}} = (3 - 400) \times 10^{30}$ erg.
Outliers are likely to be caused due to small errors in the estimate
of the warm-target temperature and the related low-energy cutoff, 
which are hugely amplified in the resulting nonthermal energies.
If we remove the outliers (in excess of $\gapprox 3$ standard
deviations in the tails of the Gaussian distribution in Fig.~2b), 
we obtain 55 events with a ratio of
\begin{equation}
	E_{\mathrm{nt,e}}/E_{\mathrm{mag}} = 0.51 \pm 0.17 \ ,
\end{equation} 
with a much smaller standard deviation factor of $\sigma=3.2$ 
as shown in Fig.~3b. 

In the previous study on the global flare energetics by 
Emslie et al.~(2012), {\it RHESSI} data have been used also, but the
warm-target model did not exist yet, since it was derived later
(Kontar et al.~2015), but is currently considered to be the best 
physical model for estimating a lower limit of the low-energy cutoff 
of nonthermal electrons (Paper III). For the temperature in the warm-target
model we used for all events the same mean value of $T_e=8.6$ MK,
which was obtained from the emission measure weighted DEMs, averaged 
during the entire flare durations, and averaged from all 
analyzed flare events.
In comparison, the low-energy cutoff value 
$e_c \approx 20$ keV of Emslie et al.~(2012), based on
the largest value that still gave an acceptable fit 
(reduced $\chi^2 \approx 1$), represents an upper limit,
while our value of $e_c \approx 6$ keV appears to be rather 
a lower limit.  The resulting
mean energy ratio of the nonthermal electron energy to the dissipated
magnetic energy was found to be 
$E_{\mathrm{nt,e}}/E_{\mathrm{mag}} = 0.03 \pm 0.02$ in Emslie et al.~(2012)
with a standard deviation factor of $\sigma=2.3$ for 
26 events (Fig.~4b). Thus the efficiency of particle acceleration
was found to be substantially lower in Emslie et al.~(2012), 
by a factor of $\approx 16$, compared with our value of 
$E_{\mathrm{nt,e}}/E_{\mathrm{mag}} 
= 0.51 \pm 0.17$ (Fig.~3b). This discrepancy appears to be the
consequence of two effects, the over-estimate of magnetic energies, 
and the adoption of upper limits for the low-energy cutoff.

\subsection{		Nonthermal Ion Energies			}

In the absence of suitable RHESSI gamma-ray data analysis we resort 
to the statistics of the earlier study by Emslie et al.~(2012), which
yields the ratios of ion energies to electron energies in
14 eruptive flare events, which we show in Fig.~5 (see also 
Table 1 and Fig.~2 in Emslie et al.~2012). The mean (logarithmic)
ratio is found to be,
\begin{equation}
	E_{\mathrm{nt,i}}/E_{\mathrm{nt,e}} = (0.34 \pm 0.50) \ ,
\end{equation}
where the standard deviation factor is $\sigma=4.5$.
Thus ions carry about a third of the energy in accelerated 
electrons (above a low-energy cutoff constrained by acceptable fits).
These flare-accelerated ion energies are
based on RHESSI measurements of the fluence in the 2.223 MeV
neutron-capture gamma-ray line (Shih 2009; Shih et al.~2009).
A caveat has to be added that the ion energies calculated in
Emslie et al.~(2012) used a low-energy cutoff of $\gapprox 1$ MeV,
and thus may represent lower limits on the ion energies and the
related ion energy ratios.

The application of the ion/electron energy ratio (Eq.~6)
to the nonthermal electron energies analyzed in our data yields
a mean (logarithmically averaged) ratio of (Fig.~3d)
\begin{equation}
	E_{\mathrm{nt,i}} / E_{\mathrm{mag}} = (0.17 \pm 0.17) \ ,
\end{equation}
which implies that about a sixth of the total dissipated magnetic
energy is converted into acceleration of ions (with energies
of $\gapprox 1$ MeV), while about half of the total magnetic energy 
(Eq.~5) goes into acceleration of electrons (above the mentioned 
low-energy cutoff of $e_c \gapprox 6$ keV).

\subsection{		Thermal Energies			}

The thermal energy in flares is mostly due to a secondary 
energy conversion process and has been quantified in Paper II.
The basic assumptions in the derivation of thermal energies are:
(1) Solar flares have a multi-thermal energy distribution;
(2) The multi-thermal energy can be calculated from the
temperature integral of the DEM distribution and a volume
estimate at the peak time of the flare;
(3) AIA data in all 6 coronal wavelengths provide a DEM in the
temperature range of $T_e \approx 0.5-20$ MK (Boerner et al.~2014),
while RHESSI is sensitive to the high-temperature tail of the 
DEM at $T_e \approx 20-40$ MK (Caspi 2010; Caspi and Lin 2010; 
Caspi et al.~2014, 2015; Ryan et al.~2014); 
(4) A suitably accurate DEM method is the spatial synthesis 
method (Aschwanden et al.~2013, 2015b),
which fits a Gaussian DEM in each spatial (macro)pixel
of AIA images in all coronal wavelengths and synthesizes the
DEM distribution by summing the partial DEMs over all (macro)pixels;
(5) The flare volume can be estimated from the geometric
relationship $V \approx A^{3/2}$, where the flare area $A$ is
measured above some suitable threshold in the emission measure
per (macro)pixel (assuming a filling factor of unity for
sub-pixel features);
(6) The thermal energy content dominates at the flare peak,
while conductive and radiative losses as well as secondary
heating episodes (indicated by subpeaks in the SXR and
EUV flux) are neglected in our analysis. The thermal energy
derived here thus represents a lower limit. 
Note that the thermal energy is calculated at the flare peak 
time (when the peak value of the total emission measure is
reached), and thus represents the peak thermal energy,
while non-thermal energies in electrons (Section 2.2) are 
calculated by time integration over the entire flare duration.

In our previous global energetics study on thermal energies (Paper II)
we were able to derive the thermal energy in 391 flare events
(of GOES M- and X-class), and find an energy range of
$E_{\mathrm{th}}=(0.15 - 215) \times 10^{30}$ erg. If we want to compare
these thermal energies with nonthermal energies, the sample reduces
to 189 events, yielding a mean (logarithmic) ratio of 
$E_{\mathrm{th}}/E_{\mathrm{nt,e}} = (0.15 \pm 0.15)$, 
with a standard deviation
factor of $\sigma = 6.5$ (Fig.~6a). Removing the outliers
by restricting the valid energy range to $E_{\mathrm{nt,e}}=(3 - 400) \times
10^{30}$ erg (see Fig.~2b), we obtain a somewhat more accurate
value for 149 events (Fig.~6b), 
\begin{equation}
	E_{\mathrm{th}} / E_{\mathrm{nt,e}} = (0.12 \pm 0.11) \ .
\end{equation}
This means that only 12\% of the nonthermal energy in electrons
is converted into heating of the flare plasma, which appears to be
a low value for the (warm) thick-target bremsstrahlung model. 
Alternative studies find that thermal and nonthermal energies
are of the same magnitude (Saint-Hilaire and Benz 2005;
Warmuth and Mann 2016a,b). 
However, since we neglected conductive and radiative losses as
well as multiple heating episodes (besides the flare peak),
the thermal energy may be grossly underestimated. In addition,
the nonthermal energy in electrons may be over-estimated
due to the lower limit of the low-energy cutoff $e_c \approx 6$ keV. 
However, since electron beam-driven chromospheric plasma heating 
is a secondary energy dissipation process, it does not affect 
the energy closure relationship (Eq.~1) of primary energy 
dissipation processes. 

Comparing the thermal energy with the available magnetic energy
we consequently find a relatively low value of (Fig.~3a) for
170 events, 
\begin{equation}
	E_{\mathrm{th}} / E_{\mathrm{mag}} = (0.08 \pm 0.13) \ .
\end{equation}
The previous study by Emslie et al.~(2012) finds an even 
lower value of $E_{\mathrm{th}} = (0.005 \pm 0.15) \ E_{\mathrm{mag}}$ 
(Fig.~4a), which
is mostly caused by the use of an iso-thermal definition of the thermal 
energy. A multi-thermal definition would yield a factor
of 14 higher values for the thermal energy (Paper II). Moreover,
no high-resolution imaging data in SXR and EUV were 
available in the study of Emslie et al.~(2012).
Even now, SXR images are available from EIS/Hinode occasionally
only, but were not used in this study.

Besides electron-beam heating of the chromosperic thick target,
non-beam heating or direct heating may also play a role 
(e.g., Sui and Holman 2003; Caspi and Lin 2010; Caspi et al.~2015). 
We derive lower
limits for the energy of direct heating processes for those flares
where the thermal energy exceeds the nonthermal energy in electrons
and ions (see Fig.~1), which yields a lower limit of 
\begin{equation}
	E_{\mathrm{dir}} / E_{\mathrm{mag}} = 
	(E_{\mathrm{th}} - E_{\mathrm{nt,e}} - E_{\mathrm{nt,i}}) / E_{\mathrm{mag}} = 
	(0.07 \pm 0.17) \ .
\end{equation}

\subsection{	Radiated Energy from Hot Flare Plasma 	}

In this section we examine the thermally radiated energy over all 
wavelengths from the hot ($>$4\,MK) coronal flare plasma.  We determined
these energies from tables of the radiative loss rate as a function of 
emission measure and temperature generated using the CHIANTI atomic 
physics database (Dere et al.~1997; Del Zanna et al.~2015) and the 
methods of Cox and Tucker (1969).  The temperatures and emission 
measures were calculated using the ratio of the GOES/XRS 0.5--4 \ang\
to 1--8 \ang\ channels (Thomas et al.~1985; White et al.~2005). 
As part of these calculations, coronal abundances (Feldman et al.~1992), 
ionization equilibria (Mazzotta et al.~1998), and a constant density 
(10$^{10}$ cm$^{-3}$) were assumed.  In addition, this methodology 
implicitly assumes that the plasma is isothermal, although this is 
not the case for the flares analyzed here and in general 
(Aschwanden et al.~2015). The isothermal 
assumption is therefore an important caveat here, but is consistent 
with previous energetics studies (Emslie et al.~2015). To ensure 
reliable results, the flare emission in both GOES/XRS channels was 
separated from the background using the Temperature and Emission 
measure-Based Background Subtraction algorithm (TEBBS; Ryan et al.~2012),
before the temperatures and emission measures were calculated.

Fig.~7a shows the coronal thermally radiated energy as a function 
of thermal energy for 389 of the 399 flares considered in this study. 
There is a correlation between flare thermal energy and radiative 
losses from the hot coronal plasma, as expected. The (logarithmic)
average ratio of radiated losses to thermal energy was found to be
\begin{equation}
		E_{rad}/E_{th} = 0.07 \pm 0.06
\end{equation}
This consistent with Emslie et al.~(2012) who found $E_{rad}/E_{th} 
= 0.17 \pm 0.15$.  
Fig.~7b compares the thermally radiated coronal losses to the 
total magnetic dissipated energy for the 171 flares common to 
this study and Aschwanden et al.~(2014). The 
average ratio was found to be
\begin{equation}
		E_{rad}/E_{mag} = 0.004 \pm 0.13
\end{equation}
From the above results it is clear that thermally radiated energy 
from the hot coronal plasma dissipates only a small fraction of the 
thermal and magnetically dissipated energies in a flare.

Although there is a positive correlation between the thermal and 
radiated energies, there is a reciprocal relationship 
in the ratio of radiated to thermal energy, as shown in Fig.~7c. 
This implies flares with larger thermal 
energy dissipate a smaller fraction of that energy via thermal 
radiation.  This is qualitatively consistent with simple hydrodynamic 
flare cooling models that predict that radiative losses and 
conductive losses are anti-correlated at higher plasma temperatures 
(e.g., Cargill et al.~1995).  No such relationship is evident from 
the results of Emslie et al.~(2012) because of their small sample 
size (38 events).

\subsection{		Bolometric Energies			}

In the largest flares, white-light emission from deep in the 
chromosphere can be observed, supposedly caused by precipitation
of nonthermal electrons and ions into the deeper chromospheric
layers (Hudson 1972). Hudson finds that the $\gapprox 5$ keV
electrons in major flares have sufficient energy to create
long-lived excess ionization in the heated chromosphere to
enhance free-free and free-bound continuum emission, visible
in broadened hydrogen Balmer and Paschen lines. Energization
to lower altitudes down to the photosphere can also be accomplished by
photo-ionization, a mechanism termed radiative backwarming
(Hudson 1972).

Kretzschmar (2011) recently demonstrated that white-light continuum
is the major contributor to the total radiated energy in most
flares, where the continuum is consistent with a blackbody
spectrum at $\approx 9000$ K. From a set of 2100 superimposed
C- to X-class flares, Kretzschmar (2011; Table 1) 
calculated the total solar
irradiance (TSI), which can be characterized by a scaling law
relationship between the bolometric energy $E_{\mathrm{bol}}$ (in erg) 
and the GOES 1-8 \ang\ SXR flux $F_{\mathrm{SXR}}$ in units of 
W m$^{-2}$ (Fig.~8a; top panel),
\begin{equation}
	{ E_{\mathrm{bol}} \over 10^{30} \ {\rm [erg]} } 
	\approx \left( {F_{\mathrm{SXR}} \over 
	2.0 \times 10^{-6}\ {\rm W}\ {\rm m}^{-2}} \right)^{0.78} \ .
\end{equation}
If we apply this empirical scaling law to the GOES fluxes and
thermal energy $E_{\mathrm{th}}$ from AIA data analyzed in Paper II, 
we obtain an energy ratio $E_{\mathrm{bol}}/E_{\mathrm{th}}$ of almost unity 
(Fig.~8b),
\begin{equation}
	E_{\mathrm{bol}} / E_{\mathrm{th}} = (1.14 \pm 0.05) \ ,
\end{equation}
and thus the bolometric energy matches
almost the thermal energy contained in the coronal flare plasma
observed in SXR and EUV. The total flare irradiance 
was found to exceed SXR emission by far. 
Woods et al.~(2004) report that 19\% of the total emission comes from the
XUV range (0-27 nm), which implies that SXR emission amounts to less than
a fifth of the total emission. Both Woods et al.~(2006)
and Kretzschmar (2011) report that only 1\% of the total bolometric
luminosity is radiated in the GOES SXR range (1-8 A).
Since both the bolometric and
the thermal energy are secondary or tertiary energy conversions
in the flare process (Fig.~1), they do not matter to the primary energy
closure (Eq.~1) investigated here, but allow us to set limits
on each energy conversion process. 

\subsection{		CME Energies				}

Almost all large flares are accompanied by a CME, and even most
mid-sized flares are associated with a CME, down to the GOES 
C-class level (Andrews 2003).
The total energy of a CME can be calculated either from the
white-light polarized brightness in coronagraph images, or from
the EUV dimming in the CME footpoint area. We used the second
method to calculate a statistical sample of CME energies using
AIA data (Paper IV). The main assumptions in our analysis are:
(1) A flare-associated dimming of the total emission measure 
observed in EUV and SXR indicates a mass loss in the
flare area, which constitutes the existence of a CME event
(Aschwanden et al.~2009; Mason et al.~2014, 2016);
(2) The DEM obtained from AIA in the temperature range of
$T_e \approx 0.5-20$ MK largely rules out that the observed
dimming is a temperature (heating or cooling) effect, because
the particle number in a CME is approximately conserved when a 
DEM is integrated over the full coronal temperature range;
(3) The EUV dimming profile is expected to drop from a higher
preflare level after the CME starts in the impulsive flare
phase, but an initial compression (or implosion) process
can produce an initial increase in the EUV total emission
measure before the EUV dimming sets in;
(4) The spatial synthesis method (Aschwanden et al.~2013),
which fits a Gaussian DEM in each spatial (macro)pixel
of AIA images in all coronal wavelengths and synthesizes the
DEM distribution by summing the partial DEMs over all 
(macro)pixels, provides a suitable method to calculate the
evolution of the total emission measure;
(5) The temporal evolution of a CME in the EUV dimming phase
can be modeled with a radial adiabatic expansion process,
which accelerates the CME and produces a rarefaction of the
density inside the CME leading edge envelope;
(6) The volume of a CME can be quantified by the footpoint
or EUV dimming area and the vertical density scale height 
of a hydrostatically stratified corona initially, and with a
reciprocal relationship between the density and volume during
the subsequent adiabatic expansion phase.  
(7) The total energy of a CMEs consists of the kinetic energy
and the gravitational potential energy to lift a CME from the
solar surface to infinity. The pressure in CMEs is modeled
with adiabatic expansion models, and thus neglects temperature 
changes during the initial expansion phase of the CME (Paper IV);
(8) A subset of non-eruptive flares, called {\sl confined
flares}, does not produce a CME, in which case our calculation
of a CME energy corresponds to the energy that goes into the 
adiabatic expansion up to a finite altitude limit where the
eruption stalls. 

In our previous global energetics study on CME energies (Paper IV)
we were able to derive the CME energy in all 399 flare events
(of GOES M- and X-class), and find an energy range of
$E_{\mathrm{CME}}=(0.25 - 1000) \times 10^{30}$ erg. 
Removing a few outliers with the highest energies 
that show an excess of $\gapprox 2$ standard deviations 
in the upper tail of a statistical random 
distribution (Fig.~2d), we obtain an improved valid range of
$E_{\mathrm{CME}} = (0.25 - 100) \times 10^{30}$ erg for the 
remaining 386 events (or 97\% of the entire data set). 

Comparing the CME energies with those events where the magnetic 
energy could be calculated, we find 157 events with a mean 
(logarithmic) energy ratio of (Fig.~3e),
\begin{equation}
        E_{\mathrm{CME}} / E_{\mathrm{mag}} = (0.07 \pm 0.14)  \ .
\end{equation}
In complex CME events with multiple convolved EUV dimming phases,
in particular for SEP events (Table 1), the CME speed, and thus 
the kinetic CME energy is likely to be substantially underestimated
with the EUV dimming method, in which case (Fig.~3e) we substitute the
AIA-inferred CME values with the Large Angle and Spectrometric
Coronagraph Experiment (LASCO)-inferred white-light values
(whenever the LASCO CME energy is larger than the AIA CME energy).
The AIA CME energies are shown in Fig.~3e, and their comparison
with LASCO CME energies is shown in Fig.~18 of Paper IV. The LASCO
CME energies were found to be larger than the AIA values in 42\%.
On the other hand, LASCO underestimates the CME energy also,
in particular for halo CMEs, because the occulted material is missing
and because the projected speed is a lower limit to the true 3-D speed. 
In other words, both the LASCO and the AIA method provide lower
limits of CME energies, which is the reason why we use the higher value
of the two lower limits as the best estimate of CME 
energies here (Fig.~3e). Therefore, less energy goes into 
the creation of a CMEs (Eq.~15) than what goes into the acceleration 
of nonthermal particles (Eq.~5).

The previous study by Emslie et al.~(2012) finds about a factor
of two higher mean value of $E_{\mathrm{CME}}/E_{\mathrm{mag}} 
= (0.19 \pm 0.20)$ (Fig.~4e). The main reason for this difference
in the CME energy is that LASCO data (used in Emslie et al.~2012)
yield a systematically higher leading-edge velocity than the
bulk plasma velocity determined with AIA (used here mostly), which
enters the CME kinetic energy with a nonlinear (square) dependence. 
Another reason is that the convolution bias in complex events tends
to produce lower limits of CME speeds (Paper IV, Section 3.1).

\subsection{		SEP Energies				}

It is generally believed that at least two processes accelerate 
particles in the solar flare and associated CME eruptions. 
First, as discussed above, magnetic 
reconnection processes in solar flares release energy that rapidly 
accelerates ions and electrons, most of which interact in the solar 
atmosphere to produce X-rays, gamma rays, and longer-wavelength 
radiation. Some fraction of these ``flare-accelerated'' particles 
can also escape into the interplanetary medium, where they can be 
identified by their composition (e.g., Mason et al.~2004). Secondly, 
the shock wave produced by a very fast CME can accelerate electrons 
to $>$100 MeV and ions to GeV/nucleon energies.  If the shock wave 
is sufficiently broad it can accelerate SEPs on field lines covering 
$\approx 180^\circ$. Aided by pitch-angle scattering and co-rotation, 
SEPs are occasionally observed over $360^\circ$ in longitude from a 
single eruption.  With a single-point measurement it is difficult to 
determine the total SEP energy content of SEPs without assumptions 
about how SEP fluences vary with longitude and latitude.  

Fortunately, during the onset of the solar-cycle 24 maximum covered 
by this study NASA's two Solar Terrestrial Relations Observatory 
(STEREO) spacecraft, STEREO-B (STB) and STEREO-A (STA), 
moved in their $\approx 1$-AU orbits from 
$\approx 70^\circ$ east (STB) and $70^\circ$ west (STA) of Earth 
to approximately $\pm 150^\circ$, making it possible to sample SEP 
particle fluences, composition, and energy spectra at two distant 
spacecraft as well as near-Earth spacecraft. This section focuses 
on those solar events where SEP energy spectra could be measured 
with the two STEREOs as well as with the near-Earth Advanced 
Composition Explorer (ACE), the Solar and Heliospheric Observatory 
(SOHO), and GOES spacecraft.
We are confident that the 3-spacecraft events reported on here are 
dominated by CME-shock-related and not flare-related SEPs.

It was often a significant challenge to correctly associate the 
SEPs observed at three well-separated locations with a specific 
flare/CME event, especially during periods when several M and X-class 
flares occurred per day. This process was aided by CME and solar 
radio-burst data, and by measurements of the interplanetary shocks 
associated with the CME eruption. For the front-side flare events 
considered here, the near-Earth and STEREO-B spacecraft are more 
likely to detect the associated SEPs than STEREO-A, because SEPs 
generally follow the Parker spiral of the interplanetary magnetic 
field lines to the east.  

Measuring the SEP fluence over a wide energy interval often 
necessitates subtracting background from an earlier event or 
extrapolating the decay of the event in question if it becomes 
buried by a new event. Sometimes many flare/CME events occur 
on the same day and it is impossible to separate individual 
SEP events as they blend together at 1 AU. Also, some flares
have no detectable SEP events. As a result, there 
was a limited sample of events where we could obtain clean 
energy spectra at all three locations. 

Lario et al. (2006, 2013) fit Gaussian distributions to multi-spacecraft 
measurements of SEP peak intensities and fluences, using
two Helios spacecraft and IMP-8 data. They also fit the radial 
dependence of SEP intensities and fluences. Gaussians were fitted to the 
3 longitudinal points of ten 3-spacecraft events from 2010-2014
analyzed here (Table 1). 
We assumed that latitude differences can also described by a Gaussian 
with the same spread as that for longitude.

To estimate the SEP energy content requires spectra over a broad 
energy range. As in the study by Emslie et al.~(2012), these 
spectral fits were extrapolated down to 0.03 MeV and up to 300 MeV 
to estimate the total MeV cm$^{-2}$ due to protons escaping through 1 AU 
at this location. We followed earlier studies 
(Mewaldt et al.~2004; 2008a,b; Emslie et al.~2012), 
which showed that protons typically make up 
$\approx 75\%$ of the SEP energy content and added an additional 
$25\%$ to account for electrons, He, and heavier ions. 

The measured SEP pitch-angle distributions indicate that most SEPs 
observed at 1 AU have undergone pitch-angle scattering in the 
turbulent interplanetary magnetic field (IMF), which also implies 
that they are likely to cross 1 AU multiple times, increasing their 
probability of detection.  In addition, protons gradually lose 
energy in the scattering process. These effects were corrected 
by using simulations of Chollet et al.~(2010), who considered 
a range of radially dependent scattering mean free paths. 
Chollet et al.~(2010) found this correction to be reasonably 
independent of the assumed scattering mean free path.

The results of this fitting procedure are summarized in Table 1. 
There appears to be a clustering of events with SEP/CME energy ratios 
of a few percents.  The maximum intensity of the fits is at 
$\approx 40^\circ$ W, almost midway between Earth and STA, so 
the peak intensity is not well constrained. The logarithmic mean 
of the Gaussian widths is $\approx 43^\circ$; similar widths were 
obtained by Lario et al.~(2006, 2013) and Richardson et al.~(2014) 
who fit multi-point measurements of SEP peak intensities for 
larger event samples. The SEP/CME energy ratio that we obtain 
is consistent with that obtained by Emslie et al. (2012) during 
solar cycle 23.  

The energy range of the ten SEP events listed in Table 1 extends
over $E_{\mathrm{SEP}}=(1.3 - 68) \times 10^{30}$ erg. 
If SEP events are
accelerated in CME-driven shocks, they should not exceed the
total CME energy. Indeed we find a ratio (Fig.~9a) of
\begin{equation}
	E_{\mathrm{SEP}}/E_{\mathrm{CME}} = 0.03 \pm 0.45 \ ,
\end{equation} 
which is comparable with the previous result of Emslie et al. (2012),
i.e., $E_{\mathrm{SEP}}/E_{\mathrm{CME}} \approx 0.04$.

Comparing the SEP energy with the total dissipated magnetic energy
of the flare, we have only 4 events available,
which yields a large uncertainty (Fig.~9b), 
\begin{equation}
	E_{\mathrm{SEP}} / E_{\mathrm{mag}} = (0.10 \pm 1.64) \ .
\end{equation}
The low ratio is consistent with our notion of CME-driven acceleration
leading to SEP events being a secondary energy conversion process
(Fig.~1). The first step supplies the generation of a CME,
while the second step drives particle acceleration in CME-driven
shocks. In particular, the low ratio confirms that the magnetic 
free energy in the flare region is sufficient to explain the
energetics of SEP particles, regardless of whether they are accelerated
in the coronal flare region or in interplanetary shocks. 

\section{	ENERGY CLOSURE 					}

After we discussed the calculations of the various forms of energy 
that occur in flares and CMEs, we are now in the position to test
the energy closure. We evaluate the energy closure for primary
energy dissipation only (Fig.~1), by adding up the nonthermal energy
in particles, $E_{\mathrm{nt}}$, the CME energy, $E_{\mathrm{CME}}$,
and the direct heating energy $E_{\mathrm{dir}}$,
which constitutes the right-hand part of Eq.~(1), and which we 
denote as the sum, $E_{\mathrm{sum}}$,
\begin{equation}
	E_{\mathrm{sum}} = 
	( E_{\mathrm{nt}} + 
	  E_{\mathrm{dir}} + E_{\mathrm{CME}} ) = 
	( E_{\mathrm{nt,e}} + E_{\mathrm{nt,i}} + 
	  E_{\mathrm{dir}} + 
          E_{\mathrm{CME,kin}} + E_{\mathrm{CME,grav}} ) \ .
\end{equation}
The ratio of these energy sum values $E_{\mathrm{sum}}$ and the dissipated
magnetic energy $E_{\mathrm{mag}}$ is shown in Fig.~2e for all 76 events
with overlapping magnetic, nonthermal, and CME data, yielding
a ratio of $E_{\mathrm{sum}} / E_{\mathrm{mag}} = 0.99 \pm 0.19$.
If we remove the outliers, as indicated by the excessive values in
the tails of log-normal Gaussian distributions (Fig.~2b and 2d),
we have a smaller sample with 54 events, but obtain a somewhat 
more accurate ratio of (Fig.~3f),
\begin{equation}
	E_{\mathrm{sum}} / E_{\mathrm{mag}} = (0.87 \pm 0.18) \ .
\end{equation}
The standard deviation of the ratio is a factor of $\sigma = 4.6$
(Fig.~2e), which shrinks after the elimination of outliers (Fig.~3f)
to a more accurate value of $\sigma = 3.2$. 
Thus we obtain an almost identical ratio with or without removal
of outliers, but a narrower standard deviation.
Our chief result is that we obtain, in the statistical average, 
energy closure for magnetic energy dissipation in flares by
$87\%$ , with an error of $\pm 18\%$ that includes the ideal
value of $100\%$ for perfect closure. This key result,
demonstrated here for the first time, is visualized
in form of a pie chart in Fig.~10 (right-hand side).

For comparison we show also the energy closure applied to the
study of Emslie et al.~(2012), as illustrated in Fig.~10 
(left-hand side).
That study has a smaller statistics
with 37 events, which provides only 8 events with overlapping
magnetic, nonthermal, and CME data, and
exhibits incomplete energy closure with a value of
$E_{\mathrm{sum}} / E_{\mathrm{mag}} = (0.25 \pm 0.24)$ (Fig.~4f).
We conclude that the overestimate of the magnetic energy 
$E_{\mathrm{mag}}$ and
the overestimate of the low-energy cutoff $e_c$ in the nonthermal energy
$E_{\mathrm{th}}$ are mostly responsible for the lack of energy closure
in the previous study of Emslie et al.~(2012).

The pie chart shown in Fig.~10 depicts that the nonthermal 
electron energy dissipates the largest fraction of magnetic 
energy, the ions dissipate the second-largest energy fraction, 
while the CMEs and direct heating require substantially less energy.
The agreement between the energy sum and the magnetic dissipated
energy varies by a standard deviation factor of $\sigma=4.6$
(Fig.~2e), which quantifies the accuracy of energy closure
that we currently are able to deduce. Since the standard deviation
of electron energies amounts to a factor of $\sigma=7.7$ (Fig.~2a),
being the largest among all forms of energies, we suspect that the
low-energy cutoff $e_c$ contains the largest uncertainty of all
parameters measured here (although we do not know the uncertainty
in the ion energy cutoff). In the largest analyzed flares,
where the electron energy was found to be systematically higher
than the dissipated magnetic energy (Paper III; Fig.~7 therein),
our method obviously over-estimates the energy in nonthermal
electrons.

Of course, there are a number of caveats,
such as the lack of energy estimates for direct heating (for which
no quantitative analysis method exists), or the lack of energy estimates
in accelerated ions (which can only be obtained in flares with
detectable gamma-ray lines and may be feasible in about 5-10
events in our data set; Albert Shih, private communication 2016).

\section{	DISCUSSION 					}

Quantifying the amount of energies in the various dynamical processes
that take place during a solar flare and CME allows us to 
to discuss which energy conversion processes are possible and which
ones are ruled out, based on the available energy. 

\subsection{	The Warm-Target Low-Energy Cutoff 		}

We found that the nonthermal energy in electrons accelerated during
a flare dissipates the largest amount of magnetic energy. This implies
that the low-energy cutoff energy $e_c$ is the most critical parameter
in the calculation of the energy budget of flares, because of the 
highly nonlinear dependence of the nonthermal energy on this
parameter. We explicitly show this functional dependence 
$E_{\mathrm{nt,e}}(e_c)$ in Fig.~11, for four different power law slopes
of the hard X-ray photon spectrum ($\gamma =4-7$), corresponding
to power law slopes $\delta=\gamma + 1$ with a range of 
$\delta = 5-8$ of the electron injection
spectrum, according to the thick-target model (Brown 1971).
From the diagram in Fig.~11 it is clear that the nonthermal energy 
varies by one to three orders of magnitude, depending on whether a 
low-energy cutoff of $e_c=6$ keV or $e_c=20$ keV is chosen.
The warm-target model of Kontar et al.~(2015) offers a new method
to constrain this low-energy cutoff, i.e., $e_c = \delta k_B T_e$,
but a reliable method to choose the correct temperature for the
warm-target has not been established yet. This may be a difficult
task, since the relevant temperature may be a mixture of cool
pre-flare plasma and hot upflowing evaporating flare plasma.
As a first attempt we used the DEM peak temperatures evaluated
from AIA data, which yield a mean temperature of $T_e=8.6$ MK
or $k_B T_e = 0.74$ keV (Paper III). This yields then a low-energy
cutoff of $e_c = \delta k_B T_e \approx 3.7-5.9$ keV for
$\delta = 5-8$. Such low values of the low-energy cutoff
have dramatic consequences.

Since the warm-target offers a physical model of the low-energy cutoff,
for which we infer a typical value of $e_c \approx
6$ keV (based on a mean temperature of $T_e=8.6$ MK in flaring 
active regions), we obtain consequently one to three orders of magnitude
higher nonthermal energies in electrons, which constrains 
a lower limit of the energy cutoff, or an upper limit for 
nonthermal electron energies. Because of the highly nonlinear 
dependence of the nonthermal energy on the low-energy cutoff, 
it produces the largest uncertainty in the nonthermal energy.

The relative energy partition of 
nonthermal electrons is the largest difference to the study of
Emslie et al.~(2012), which is explained by the highly nonlinear
scaling behavior of the low-energy cutoff (see Fig.~11 for estimates
of the relative change in the energy partition). It dominates all 
other energetics, is mainly responsible for the energy closure, 
and together with the lower CME energies it reverses the
flare-CME energy partition derived by Emslie et al.~(2012), and in
addition completely dominates over the thermal flare energy,
in contrast to the results of Saint-Hilaire and Benz (2005) and 
Warmuth and Mann (2016a, b). It is clear that these new contrasting
results mostly occur due to the adoption of a relatively low 
energy cutoff imposed by the warm-target model. 
For instance, the nonthermal energy for event \#12 (Table 2)
exceeds the dissipated magnetic energy substantially and is
likely to be over-estimated due to a large error in the
low-energy cutoff. Hence, the assumption of the warm-target 
temperature, the measurement of a representative temperature 
distribution in the inhomogeneous flare plasma, and its 
variation from flare to flare, are subject to large uncertainties,
and thus add a significant caveat to our energy closure tests.
In order to minimize uncertainties of the assumed warm-target
temperature, we used a mean value of $T_e=8.6$ MK that was
obtained from the emission measure weighted DEMs, averaged 
during the entire flare durations, and averaged from all 
analyzed flare events.

\subsection{	Sufficiency of the Thick-Target Model   }

In the classical (cold) thick-target bremsstrahlung model (Brown 1971), 
nonthermal electrons precipitate from the coronal acceleration
site along the magnetic field lines towards the chromosphere,
heat up the plasma in the upper chromosphere and drive upflows
of heated plasma, a process that is called {\sl chromospheric
evaporation}. In this scenario, all nonthermal energy of the
precipitating electrons is converted into the thermal energy 
of the evaporating plasma. Therefore, in the absence of any
other heating mechanism, we expect the inequality,
\begin{equation}
	E_{\mathrm{th}} \le E_{\mathrm{nt}} = 
	( E_{\mathrm{nt,e}} + E_{\mathrm{nt,i}} ) \ .
\end{equation}
We discussed this inequality in Section 2.4 and showed
that virtually all flares have a thermal energy that is
substantially less than the nonthermal energy in electrons
(Fig.~6), after removal of statistical outliers. This result
confirms that the thick-target bremsstrahlung model is
sufficient to explain the observed thermal plasma in flares. 

\subsection{	Secondary Energy Dissipation Processes	}

While we discussed only the primary energy dissipation processes
in Section 3, we may also consider secondary energy dissipation
processes for the energy balance, 
which includes the generation of thermal energy,
bolometric energy, and radiative energies in flares, as depicted
in the diagram of Fig.~1. Ignoring the CME-related energies for
the moment, most of the non-thermal energy in accelerated
electrons and ions, as well as direct heating, is expected
to contribute to the thermal energy $E_{th}$, based on the
thick-target model and the Neupert effect, where precipitating
electrons heat up the coronal warm-target regions and the upper
chromosphere by the so-called chromospheric evaporation process.
Interestingly, however, we measure thermal energies that amount
to 12\% of the non-thermal energies only (Fig.~6b). Does this
imply a low efficiency of the thick-target model? There are
essentially two possibilities: either the non-thermal energy 
in electrons is over-estimated (most likely because of the 
relatively low cutoff energy of 6 keV), or the thermal energy is
under-estimated (mostly because we calculate the thermal
energy at the flare peak time only).

On the other side, one would expect that the bolometric energy
should constitute at least a major fraction of the non-thermal
energy in electrons and ions, as well as the resulting thermal
energy, manifested by white-light emission in deeper chromospheric 
layers due to locally enhanced ionization. Indeed we do find 
that the bolometric energy equates to the thermal energy in the
statistical average ($E_{bol}/E_{th}=1.14\pm0.05$, Fig.~8b),
but there is a discrepancy that the bolometric energy does not
match the non-thermal energy in electrons, estimated to be
$E_{bol}/E_{nt,e}=0.07/0.51 \approx 0.14$ (based on  
$E_{bol}/E_{mag}=0.07\pm0.10$ and   
$E_{nt,e}/E_{mag}=0.51\pm0.17$; Table 3). 
A good result of this estimates is that the bolometric energy
approximately matches the thermal energy, consistent with
other findings for very large flares, where two independent 
methods of determining $E_{bol}$ give a similar balance,
using single events from SORCE and event ensembles from SOHO/VIRGO
(Warmuth and Mann 2016a,b). We suspect that our method
may over-estimate the nonthermal energy and thus yields an upper
limit on the energy in non-thermal electrons, complementary 
to the lower limits (or under-estimates) of other earlier
studies (Emslie et al.~2012).

\subsection{	Magnetic Reconnection Models  		}

Our result of energy closure (Eqs.~18, 19) corroborates the 
conjecture that a flare with (or without) CME is of magnetic origin.
Stating this result the other way round, we conclude that no
other (than magnetic) energy sources are needed to produce a flare 
or to expel a CME. As we mentioned in Section 2.1, the dissipated
magnetic energy was calculated from the twist of helical field 
lines in the flaring active region that is relaxed during a flare
and leads to a lower (magnetic) energy state. We may ask what kind
of magnetic processes are consistent with this scenario? Magnetic
reconnection is most generally defined by a mutual exchange of
the connectivity between oppositely polarized magnetic charges.
In the case of solar flares, the magnetic charges are buried
below the photospheric surface, while the coronal configuration
of the magnetic field can be bipolar, tripolar, or quadrupolar.
A magnetic reconnection process needs to be triggered by a
magnetic instability, but evolves then from a higher to a lower
energy state. This is reflected in our finding that the free
energy reduces from a higher value at flare start to a lower
value at flare end (Paper I). However, a puzzling observation
is that often an increase of the free energy is observed 
immediately before flare start (Aschwanden, Xu, and Jing 2014, 
Paper I), which is not predicted by 
magnetic reconnection process. Such a feature could be produced
by temporary compression or an implosion process, but is poorly
understood at this point. Nevertheless, our result on the energy
closure strongly confirms the role of magnetic reconnection models,
and could not be explained in terms of any non-magnetic process
(such as by acoustic waves or hydrodynamic turbulence).

\subsection{	The Acceleration Efficiency 		}

Our result on the nonthermal energy in electrons amounting
to approximately half of the dissipated magnetic energy (Eq.~5)
implies a highly efficient accelerator, at least for electrons.
From the statistical result of
 $E_{\mathrm{nt,e}}/E_{\mathrm{mag}} \approx 0.5$ (Eq.~5) obtained
from our measurements we can estimate the required electron
densities and magnetic fields in the acceleration region.
The electron spectrum falls off steeply with energy, so that
the mean kinetic energy of accelerated electrons is essentially
given by the low-energy cutoff $e_c = (1/2) m_e v^2 \approx
6$ keV $\approx 10^{-8}$ erg. Thus we obtain the total kinetic 
energy of all accelerated electrons by multiplying the kinetic energy
of a single (nonthermal) electron with the density $n_{acc}$
of accelerated electrons and volume $V$ of the acceleration region,
\begin{equation}
	E_{\mathrm{nt,e}} = ({1 \over 2} m_e v^2) \ n_{acc} \ V 
	         \approx e_c \ n_{acc} \ V \ .
\end{equation}
On the other side, the total free magnetic energy is given by the
volume integral,
\begin{equation}
	E_{\mathrm{mag}} = \left( {B_\varphi^2 \over 8 \pi } \right) \ V \ .
\end{equation}
Setting the energy ratio to the observed value,  
$E_{\mathrm{nt,e}} = E_{\mathrm{mag}}$ (Eq.~5) yields then
for the acceleration efficiency $q_{\mathrm{acc}}$,
\begin{equation}
	q_{\mathrm{acc}} = 
	\left( {E_{\mathrm{nt,e}} \over E_{\mathrm{mag}}} \right) = 0.5 \times 
	\left( {e_c      \over 6\ {\rm keV}} \right) 
	\left( {n_{acc}  \over 10^9\ {\rm cm}^{-3} } \right) 
	\left( {B_{\varphi} \over 22\ {\rm G}} \right)^{-2} \ .
\end{equation}
Thus for a moderate potential field of $B_p \approx 100$ G and
a twisted perpendicular component of $B_\varphi \approx 22$ G,
which corresponds to a twist angle of 
$\alpha = \arctan{(B_{\varphi} / B_{p})} \approx 12^\circ$,
we can explain electron acceleration above a low-energy cutoff
of 6 keV. If we insert the measured acceleration efficiency of
$q_{\mathrm{acc}} \approx 0.5$ and the associated low-energy cutoff 
value of $e_c = 6$ keV, we obtain a direct relationship between
the mean azimuthal magnetic field $B_{\varphi}$ and the mean electron
density $n_{acc}$,
\begin{equation}
	\left( {n_{acc}      \over 10^9\ {\rm cm}^{-3} } \right) 
	\approx \left( {B_{\varphi} \over 22\ {\rm G}} \right)^{2} \ ,
\end{equation}
which provides us another testable relationship in the flaring
active region. The azimuthal field component $B_{\varphi}$ can directly 
be measured with the {\sl vertical-current approximation nonlinear 
force-free field (VCA-NLFFF)} code used in Paper I, while the
mean electron density can be obtained from the total emission
measure and flare volume as measured in Paper II. However,
the spatio-temporal flare geometry has to be deconvolved into 
single flare loops for a proper test.

\subsection{	Conductive and Radiative Energy Losses 	}

The heated solar flare plasma, which is produced by chromospheric
heating from precipitating electrons and ions (and direct heating),
and by subsequent chromospheric evaporation, loses its thermal 
energy by conductive and radiative losses in the solar corona,
according to the Neupert effect. In addition, some flare
plasma will be directly heated in the acceleration region
(e.g., Sui and Holman 2003; Caspi and Lin 2010; 
Liu et al. 2013; Caspi et al.~2015), 
for which we can estimate a lower limit for the cases 
where the thermal energy excceds the nonthermal energy. 
Since we consider the acceleration of electrons and ions as
primary energization process in our energy budget (Fig.~1), 
all subsequent heating and cooling processes are secondary 
energy conversion steps and thus are not included in the 
energy budget in order to avoid double-counting. These cooling 
processes include energy losses due to (1) thermal conduction
from the corona to the chromosphere, with a energy loss rate
$dE_{\mathrm{cond}}/dt \propto -T^{7/2}/L^2$ that tends to be
most efficient for the hottest flare plasma and the shortest 
flare loops, and
additional energy losses due to (2) radiative losses, with
a radiative cooling rate of $dE_{\mathrm{rad}}/dt \propto -n_e^2
T^{-3/2}$, being most efficient in the densest flare loops
at lower temperatures radiating in EUV. Radiative losses
in soft X-rays, calculated from the GOES fluxes, yielded a
very small contribution to the total energy budget, i.e.,
$E_{rad}/E_{mag}=0.004\pm0.130$ (Eq.~12, Section 2.5).

In principle the
total energy losses can be computed for each flare event,
but this would require to measure the time evolution of the
volumetric heating rate and conductive and radiative losses 
with proper spatio-temporal modeling, which is not attempted
in our statistical study, since radiative energy 
losses amount to a negligible fraction of the global flare
energy budget. For more details, 
the reader is referred to the study of Milligan 
et al.~(2014), where the radiated energy budget of chromospheric
plasma in a major solar flare is deduced from multi-wavelength
observations. We quote in Table 2 the energy values for
flare \#12 (2011 Feb 15, 01:46 UT; see also Fig.~3 in Paper III),
along with a condensed form of Table 3 
of Milligan et al.~(2014), which provides the energies that
are re-radiated across the visible and EUV ranges of the
solar spectrum, all being in the energy range of
$E_{\mathrm{rad}} \approx (10^{26} - 10^{30})$ erg, 
and thus are fully accounted for by the dissipated magnetic 
energies derived here.

Although one would expect in the thick-target model
that the total radiative energy loss cannot exceed the 
thermal energy, there is the possibility that continuous
energy input (by nonthermal particles and direct heating)
into the flare plasma after the flare peak can boost the
radiative energy above the thermal energy, especially in
large events. Both Emslie et al. (2012) and Warmuth and Mann 
(2016a,b) found that the radiated energy of the hot plasma 
can be slightly higher than the maximum thermal energy,
while Warmuth and Mann (2016a,b) deduced conductive losses that 
were significantly larger than the peak thermal energies.
If this is the case, radiative losses could possibly add
a non-negligible fraction to the global energy budget.

\section{	CONCLUSIONS					}

This study is the first attempt to investigate energy closure in
solar flare and CME events. All the arguments made here are based
on the various forms of energies as measured in a series of recent 
studies, which include the magnetic energy (Paper I), the thermal 
energy (Paper II), the nonthermal energy (Paper III), and CME energies 
(Paper IV). We arrive at the following conclusions:

\begin{enumerate}

\item{\underbar{Energy Closure:} From the temporal
causality that is inherent in the most commonly used physical models of 
flare and CME processes we distinguish between primary and secondary energy
dissipation processes, but test mainly the energy closure of the
primary step, which includes the dissipation of free magnetic energy
$E_{\mathrm{mag}}$ to support acceleration of particles (electrons and ions)
with a total nonthermal energy $E_{\mathrm{nt}}=E_{\mathrm{nt,e}}+E_{\mathrm{nt,p}}$,
direct heating of flare plasma $E_{\mathrm{dir}}$, 
and the simultaneous launch of a CME with a kinetic and
gravitational potential energy 
$E_{\mathrm{CME}} = E_{\mathrm{CME,kin}} + E_{\mathrm{CME,grav}}$.
Thus, the expected energy closure in the primary flare dissipation process is
the equivalence between the dissipated magnetic energy $E_{\mathrm{mag}}$ and
the sum of the first-step energy dissipation processes, 
$E_{\mathrm{sum}} = E_{\mathrm{nt}} + E_{\mathrm{dir}} + E_{\mathrm{CME}}$. 
Our chief result is the finding of
equivalence in the statistical mean, within the statistical uncertainties, 
namely $E_{\mathrm{sum}}/E_{\mathrm{mag}}= 0.87 \pm 0.18$, with a standard deviation
factor of $\sigma=3.2$ for individual flare/CME events. 
If we restrict the statistics to a subset of 76 events by eliminating outliers, 
we find an energy closure of $E_{\mathrm{sum}}/E_{\mathrm{mag}}= 
0.99\pm0.19$ (Fig.~2e).}

\item{\underbar{Energy Partition in the primary flare energy budget:}
Comparing the mean ratios of the various primary energy dissipation
processes with the dissipated magnetic energy (100\%), we find in
the statistical average, that 
51\% of the magnetic energy goes into nonthermal electrons,
17\% into nonthermal ions,
7\% into the launch of a CME,
7\% into direct heating of flare plasma,
and 18\% is the residual that may include alternative energy dissipation
processes or statistical errors. Since the analyzed data set is a complete 
sample of all flares with GOES class $\ge M1$, it is dominated by mid-size
($\gapprox$ M1.0) flares.}

\item{\underbar{The thermal/nonthermal energy ratio:}
We find a relatively low ratio of thermal to nonthermal energies, i.e.,
$E_{\mathrm{th}}/E_{\mathrm{nt,e}}=0.12 \pm 0.11$. 
This result is consistent with the 
thick-target bremsstrahlung model (Brown 1971) in the sense that
the precipitating nonthermal electrons contain sufficient energy
to heat up the upper chromosphere and to drive chromospheric evaporation
to produce the observed thermal energy in SXR and EUV.
On the other side, for an ideal thick-target model we would expect
near-equivalence of thermal and nonthermal energies. We suspect that
this low energy conversion efficiency is caused by combination of
over-estimated nonthermal energies in electrons, and
under-estimated thermal energies due to neglecting multiple (secondary) 
heating episodes and simultaneous conductive and radiative losses.}

\item{\underbar{The bolometric/thermal energy ratio:} White-light
emission appears in all large flares and is highly correlated with
the SXR flux. We find an energy ratio of $E_{\mathrm{bol}}/E_{\mathrm{th}}
=1.14 \pm 0.05$ between the bolometric energy and the thermal
energy, using the scaling law of Kretzschmar (2011) between the
bolometric luminosity and the GOES SXR flux. The 
flare-associated SXR flux is believed to be produced mostly by 
precipitating particle beams (due to the generation of hot plasma
by chromospheric evaporation), which may cause enhanced ionization
and excitation of white-light flare emission as well.} 

\item{\underbar{The SEP/CME energy ratio:} Based on the SEP analysis
of a small subset of 8 events we find a (logarithmic mean) ratio of 
$E_{\mathrm{SEP}}/E_{\mathrm{CME}}
=0.03 \pm 0.45$ between the energy in SEPs and CMEs. This
result corroborates the conjecture that SEP particles are 
primarily accelerated
by CME-driven shocks, with an acceleration efficiency in the order
of a few percents. Of course, this does not eliminate a possible
acceleration of SEPs at the coronal flare site.}

\item{\underbar{The warm-target concept} provides a physical model 
for estimating a lower limit of the low-energy cutoff $e_c$, or an 
upper limit on the nonthermal energies, which scales with the 
temperature $T_e$ of the warm-target 
plasma and the power law slope $\delta$ of the nonthermal spectrum.
Using the DEM peak temperature of a large sample of M- and
X-class flares yields a mean temperature of $T_e = 8.6$ MK and
a low-energy cutoff value of $e_c \approx 6$ keV, which is
substantially below earlier estimations of $e_c \approx 20$ keV
and produces about one to three orders of magnitude higher 
nonthermal energies. Because of the highly nonlinear dependence of 
the nonthermal energy on the low-energy cutoff, it produces the 
largest uncertainty in the nonthermal energy and in the energy 
closure relationship.}

\end{enumerate}

Energy closure constitutes a rigorous quantitative test whether
our physical models of dynamic phenomena are complete and
accurate, or whether we miss important first-order effects.
In our study on solar flares and CMEs we fortunately find
energy closure for (nonpotential) magnetic energies that supply
the creation of a flare and the launch of a CME, which
is a strong endorsement for magnetic reconnection models.
From the inequality relationships of secondary energy dissipation
processes we also find strong support for the thick-target model,
the warm-target model, flare-associated chromospheric white-light 
emission, and CME-driven shocks, but we encountered large 
uncertainties up to an order of magnitude in some of the calculated 
energies, in particular for the nonthermal energy that depends
in a highly nonlinear manner on the low-energy cutoff.
In addition, there are number of flare aspects that we do
not understand at this time, for instance: (1) The direct heating in flares
that accompanies particle acceleration; (2) The physics of
various particle transport and acceleration processes; and 
(3) The thermal evolution and shock-driven acceleration in CMEs.
Future modeling, using the powerful tool of energy closure criteria 
applied here, may further help to discriminate various physical 
flare and CME models. 

\bigskip
\acknowledgements
We acknowledge useful comments from an anonymous referee
and discussions with Gordon Emslie, Nat Gopalswamy, Ryan Milligan, 
Nariaki Nitta, Albert Shih, Manuela Temmer, Barbara Thompson, 
Astrid Veronig, Angelos Vourlidas, Alexander Warmuth, and Jie Zhang.
Contributors to the analysis of SEP events are Richard Mewaldt,
Christina Cohen, David Lario, Glenn Mason, Ian Richardson, and
Mihir Desai.
This work was partially supported by NASA contract NNX16AF92G of
the project {\sl Global Energetics of Solar Flares and CMEs},
and by NASA contract NNG04EA00C of the {\it SDO/AIA} instrument. 
Amir Caspi and Jim McTiernan were supported by NASA Grants 
NNX15AK26G and NNX14AH54G, and by NASA Contract NAS5-90833.
Christina Cohen and Richard Mewaldt were supported by NASA 
under grants NNX13A66G and subcontract 00008864NNX15AG09G 
of grant NNX15AG09G.
Daniel Ryan was supported by the NASA PostDoc program through
the Universities Space Research Association (USRA) 
and the Royal Observatory of Belgium (ROB).

%%%%%%%%%%%%%%%%%%%%%%%%%%% REFERENCES %%%%%%%%%%%%%%%%%%%%%%%%%%%%% 

\clearpage

%%%%%%%%%%%%%%%%%%%%%%%%%%% TABLE 1  %%%%%%%%%%%%%%%%%%%%%%%%%%%%%%%%%
\begin{deluxetable}{rrrrrrrr}
\tablecaption{SEP kinetic energies for selected 3-spacecraft events
from 2011-2013. The higher value of the two lower limits of CME/LASCO
(column 6) and CME/AIA energies (column 7) is used in the SEP/CME
ratio (column 8).}
\tablewidth{0pt}
\tablehead{
\colhead{\#}&
\colhead{Flare}&
\colhead{{\it GOES}}&
\colhead{Heliographic}&
\colhead{SEP kinetic}&
\colhead{CME/{\it LASCO}}&
\colhead{CME/{\it AIA}}&
\colhead{SEP/CME}\\
\colhead{}&
\colhead{Date}&
\colhead{class}&
\colhead{position}&
\colhead{energy}&
\colhead{energy}&
\colhead{energy}&
\colhead{energy ratio}\\
\colhead{}&
\colhead{}&
\colhead{}&
\colhead{}&
\colhead{$(10^{30}$ erg)}&
\colhead{$(10^{30}$ erg)}&
\colhead{$(10^{30}$ erg)}&
\colhead{}}
\startdata
	 12 	& 2011-Feb-15   & X2.2  & S21W12        &    1.3   & $>$1.6 &   161.0 & 0.008 \\
	 58	& 2011-Aug-04	& M9.3 	& N18W36	&    4.9   &  45.0  & $>$15.0 & 0.110 \\
	 74	& 2011-Sep-22	& X1.4	& N08E89	&    2.8   & 265.0  & $>$14.0 & 0.011 \\
	102	& 2011-Oct-22	& M1.3	& N27W87	&   13.6   &  22.0  & $>$17.0 & 0.620 \\
	131	& 2012-Jan-23	& M8.7 	& N33W21	&   37.3   & 413.0  & $>$19.0 & 0.090 \\
	132	& 2012-Jan-27	& X1.7	& N33W85	&   24.5   & 819.0  & $>$41.0 & 0.030 \\
	148	& 2012-Mar-07	& X1.3	& N18E29	&   67.6   & 362.0  & $>$12.0 & 0.190 \\
	169	& 2012-May-17	& M5.1	& N07W88	&    6.0   & 251.0  & $>$14.0 & 0.024 \\
	284	& 2013-May-13	& X1.7	& N11W89	&    2.0   &  61.0  & $>$11.0 & 0.033 \\
	296	& 2013-Jun-21	& M2.9	& S14E73	&    2.4   & 100.0  & $>$12.0 & 0.024 \\
                &               &       &               &          &        &         &       \\
                &Logarithmic    & mean  &               &          &        &         & 0.03$\div$3.2 \\
\enddata
\end{deluxetable}

%%%%%%%%%%%%%%%%%%%%%%%%%%% TABLE 2  %%%%%%%%%%%%%%%%%%%%%%%%%%%%%%%%%
\begin{deluxetable}{lrr}
\tablecaption{Wavelength ranges and energies of the
2011 Feb 15, 01:46 UT, {\it GOES} X2.2 flare, as derived for 
magnetic energies     (see Table 3 in Paper I),
the thermal energy    (see Table 2 in Paper II),
the nonthermal energy (see Table 1 in Paper III),
the CME energies      (see Table 3 in Paper IV), 
and radiative energies determined by Milligan et al.~(2014).}
\tablewidth{0pt}
\tablehead{
\colhead{ }&
\colhead{Wavelength range}&
\colhead{Energy}\\
\colhead{}&
\colhead{(\ang )}&
\colhead{(erg)}}
\startdata
Magnetic potential energy&6173, 94-305	&$(1065\pm 14)  \times 10^{30}$ \\
Magnetic free energy    & 6173, 94-305	&$(52  \pm 20)	\times 10^{30}$ \\
Magnetic dissipated energy&6173, 94-305	&$(120 \pm 10)	\times 10^{30}$ \\
			&		&				\\
Thermal energy 		& 94-305	&$82		\times 10^{30}$ \\
			&		&				\\
Nonthermal energy 	& 0.25-2.1      &$1100		\times 10^{30}$ \\
			&		&				\\
CME kinetic energy	& 94-305	&$124           \times 10^{30}$ \\
CME gravitational energy& 94-305	&$40            \times 10^{30}$ \\
			&		&				\\
Ly $\alpha$ line	&1170-1270	&$(1.2 \pm 0.3) \times 10^{30}$ \\
He II line		&302.9-304.9	&$(3.4 \pm 0.1) \times 10^{29}$ \\
UV continuum		&1600-1740	&$ 2.6          \times 10^{29}$ \\
C IV line + UV continuum&1464-1609	&$ 1.7          \times 10^{29}$ \\
Lyman continuum		&504-912	&$(1.8 \pm 1.0) \times 10^{29}$ \\
Ca II H line		&3967-3970	&$ 5.5          \times 10^{28}$ \\
He I continuum		&370-504	&$(3.0 \pm 0.6) \times 10^{28}$ \\
He II continuum		&200-228	&$ 1.6          \times 10^{28}$ \\
Green continuum		&5548-5552	&$ 1.5          \times 10^{26}$ \\
Red continuum		&6682-6686	&$ 1.4          \times 10^{26}$ \\
Blue continuum		&4502-4506	&$ 1.2          \times 10^{26}$ \\
\enddata
\end{deluxetable}
\clearpage

%%%%%%%%%%%%%%%%%%%%%%%%%%% TABLE 3  %%%%%%%%%%%%%%%%%%%%%%%%%%%%%%%%%
\begin{deluxetable}{lrrrr}
\tablecaption{Summary table of statistical energy ratios in flares.
The sum of primary energies includes nonthermal electrons, ions, 
direct heating, and CME (kinetic and potential) energies.}
\tablewidth{0pt}
\tablehead{
\colhead{Energy type}&
\colhead{Number of}&
\colhead{Fraction of}&
\colhead{Number of}&
\colhead{Fraction of }\\
\colhead{type}&
\colhead{flares}&
\colhead{magnetic energy}&
\colhead{flares}&
\colhead{thermal energy}}
\startdata
Free magnetic energy		& 172 &	$E_{\mathrm{mag}}/E_{\mathrm{mag}} =1.00\pm0.00  $      & \\
Nonthermal electrons	        &  55 & $E_{\mathrm{nt,e}}/E_{\mathrm{mag}}=0.51\pm0.17$ &      & \\
Nonthermal ions			&  55 & $E_{\mathrm{nt,i}}/E_{\mathrm{mag}}=0.17\pm0.17$ &      & \\
CME energy	 		& 157 & $E_{\mathrm{CME}}/E_{\mathrm{mag}} =0.07\pm0.14$ &      & \\
SEP energy	 		&   4 & $E_{\mathrm{SEP}}/E_{\mathrm{mag}} =0.10\pm1.64$ &      & \\
Direct heating 			& 106 & $E_{\mathrm{dir}}/E_{\mathrm{mag}} =0.07\pm0.17$ &      & \\
Thermal energy			& 170 & $E_{\mathrm{th}}/E_{\mathrm{mag}}  =0.08\pm0.13$ &  391 & $E_{\mathrm{th}}/E_{\mathrm{th}}=1.00\pm0.00$ \\
Radiated energy in SXR 		& 171 &	$E_{\mathrm{rad}}/E_{\mathrm{mag}} =0.004\pm0.130$& 389 & $E_{\mathrm{rad}}/E_{\mathrm{th}}=0.07\pm0.06$ \\
Bolometric energy		& 172 &	$E_{\mathrm{bol}}/E_{\mathrm{mag}} =0.07\pm0.10$ &  391 & $E_{\mathrm{bol}}/E_{\mathrm{th}}=1.14\pm0.05$ \\
Sum of primary energies         &  52 & $E_{\mathrm{sum}}/E_{\mathrm{mag}} =0.87\pm0.18$ &      &                              \\
\enddata
\end{deluxetable}
\clearpage

%%%%%%%%%%%%%%%%%%%%%%%%%%% FIGURE %%%%%%%%%%%%%%%%%%%%%%%%%%%%%%%%% 

\begin{figure}
\plotone{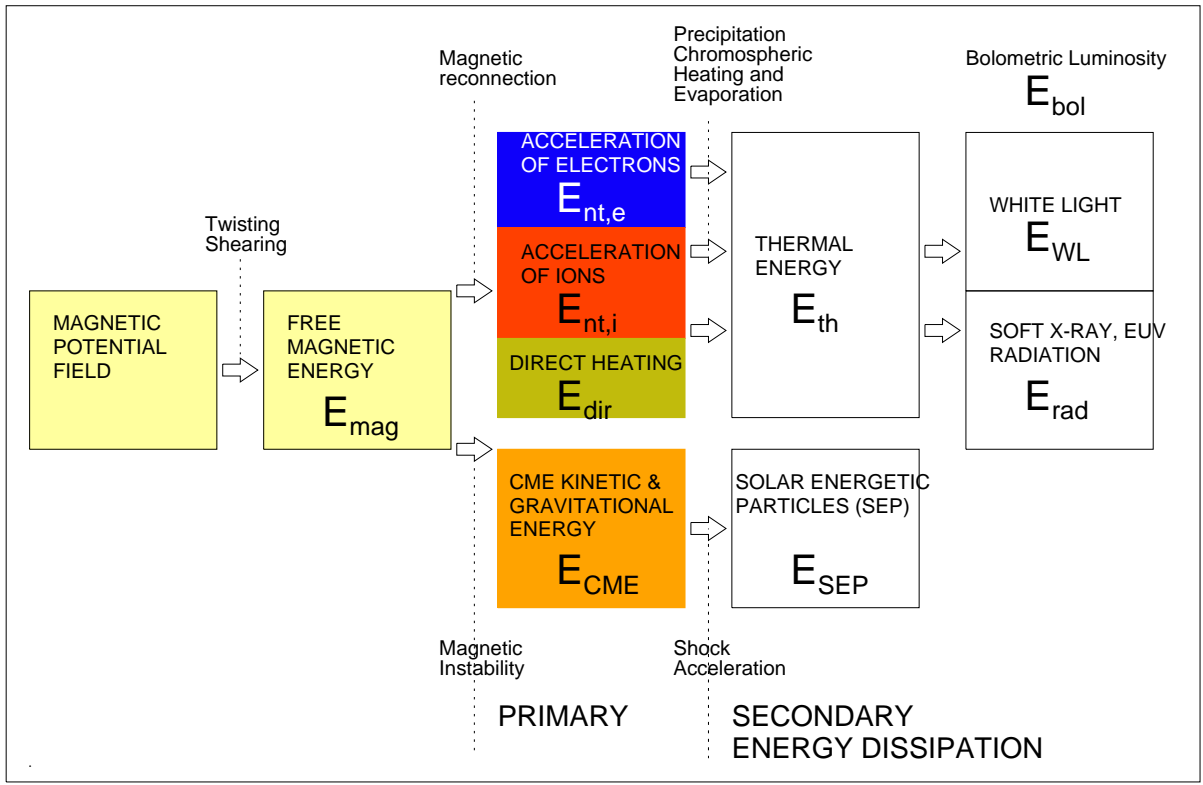}
\caption{Schematic diagram of energy input (free magnetic energy $E_{\mathrm{mag}}$),
primary energy dissipation processes (electron acceleration $E_{\mathrm{nt,e}}$, 
ion acceleration $E_{\mathrm{nt,i}}$, direct heating $E_{\mathrm{dir}}$, 
and launching of CME $E_{\mathrm{CME}}$), and secondary energy dissipation processes 
(thermal energy $E_{\mathrm{th}}$, solar energetic particles $E_{\mathrm{SEP}}$, 
and bolometric luminosity $E_{\mathrm{bol}}$, with radiative energies 
observed in white-light $E_{WL}$, soft X-rays and EUV $E_{\mathrm{rad}}$).}
\end{figure}

\begin{figure}
\plotone{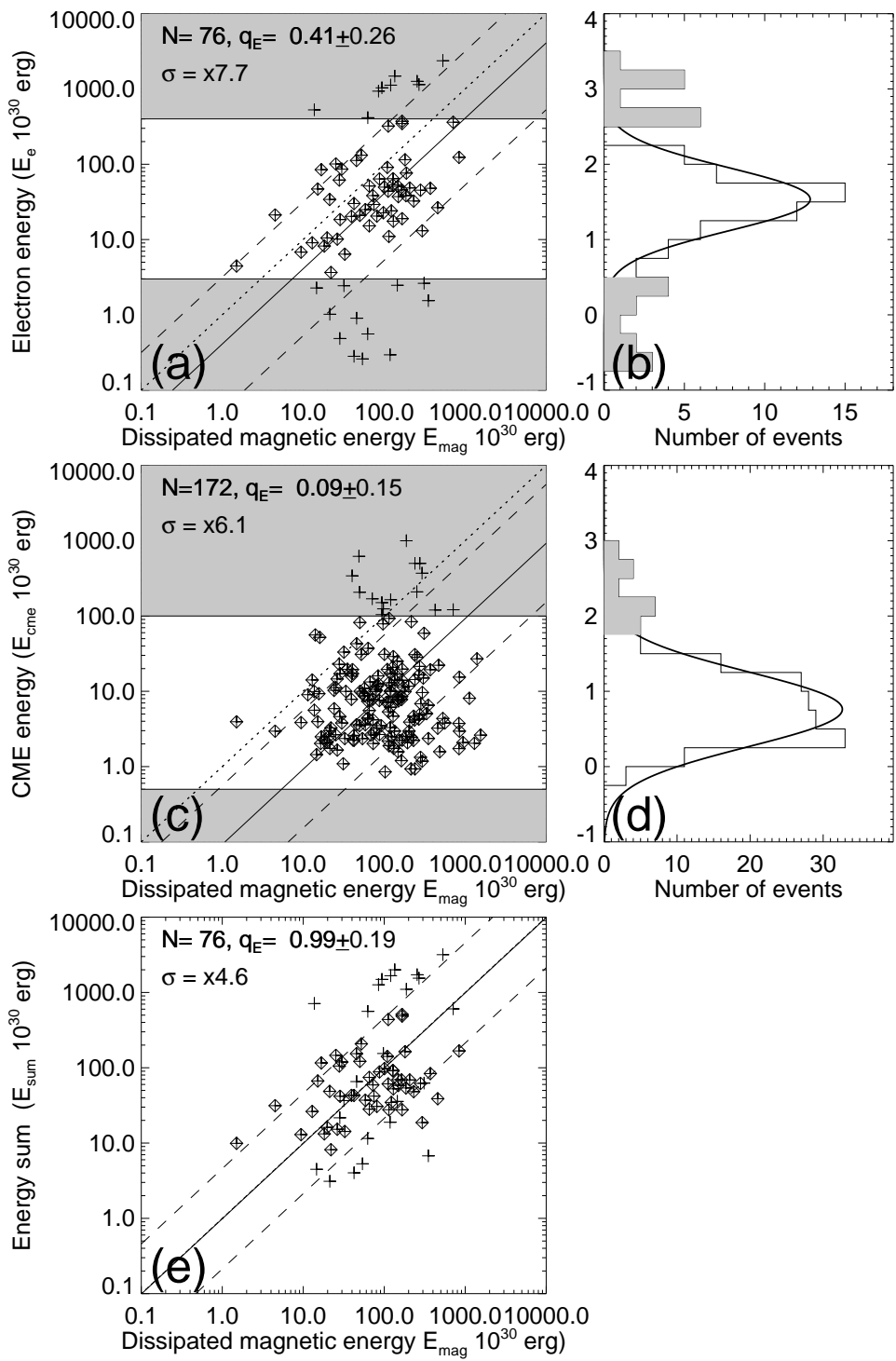}
\caption{Cross-correlation plots of the electron energy $E_{\mathrm{nt,e}}$ (a), 
the CME energy $E_{\mathrm{CME}}$ (c), and the energy sum 
$E_{\mathrm{sum}}=E_{\mathrm{nt,e}}+E_{\mathrm{nt,i}}+E_{\mathrm{dir}}+E_{\mathrm{CME}}$
with the dissipated magnetic energy $E_{\mathrm{mag}}$ (e). 
Log-normal Gaussian distributions
are fitted (b,d) to the histogrammed events and the outlier data points are marked 
with crosses and the ranges are shown with grey areas in (a,b,c,d). 
Normal data points without outliers are marked with diamonds. The mean
(logarithmic) ratios are indicated with a diagonal solid line, the standard
deviations with dashed lines, and equivalence with dotted lines.
The parameters listed in each panel include the number of events $N$,
the (logarithmic) mean energy ratio $q_E$, and the standard deviation
factor $\sigma$, as defined at the beginning of Section 2.}
\end{figure}

\begin{figure}
\plotone{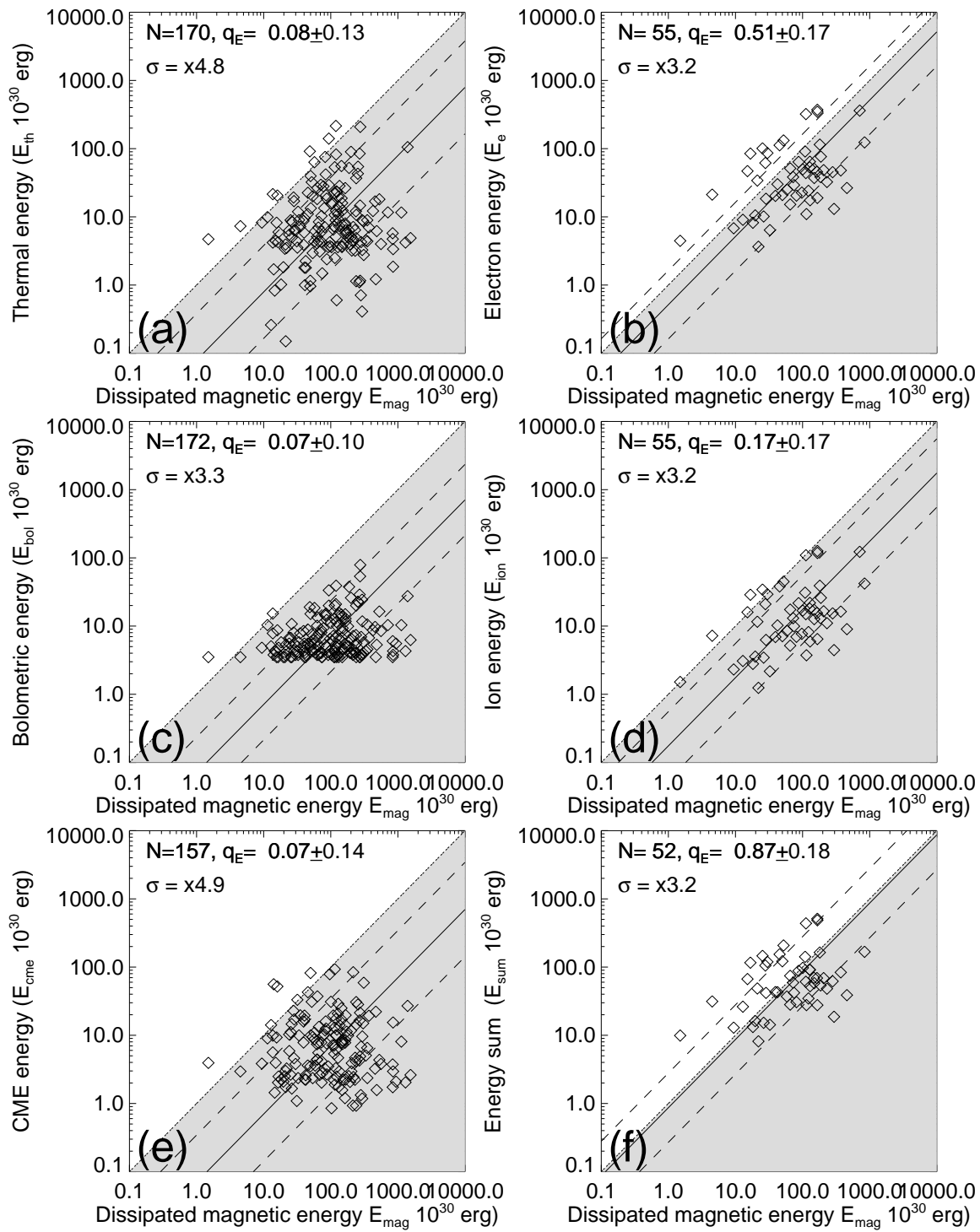}
\caption{Cross-correlation plots of 
the thermal energy $E_{\mathrm{th}}$ (a), 
the bolometric energy $E_{\mathrm{bol}}$ (c),
the CME energy $E_{\mathrm{CME}}$ (e), 
the nonthermal electron energy $E_{\mathrm{nt,e}}$ (b), 
the nonthermal ion energy $E_{\mathrm{nt,i}}$ (c), 
and the energy sum $E_{\mathrm{sum}}=E_{\mathrm{nt,e}}+E_{\mathrm{nt,i}}
+E_{\mathrm{dir}}+E_{\mathrm{CME}}$ (f),
with the dissipated magnetic energy $E_{\mathrm{mag}}$. 
Outlier events (marked with cross symbols in Fig.~2)
have been removed in this selection of datapoints.
The mean (logarithmic) ratios are indicated with a diagonal solid line, 
the standard deviations with dashed lines, and equivalence with dotted 
diagonal line bordering the grey area.}
\end{figure}

\begin{figure}
\plotone{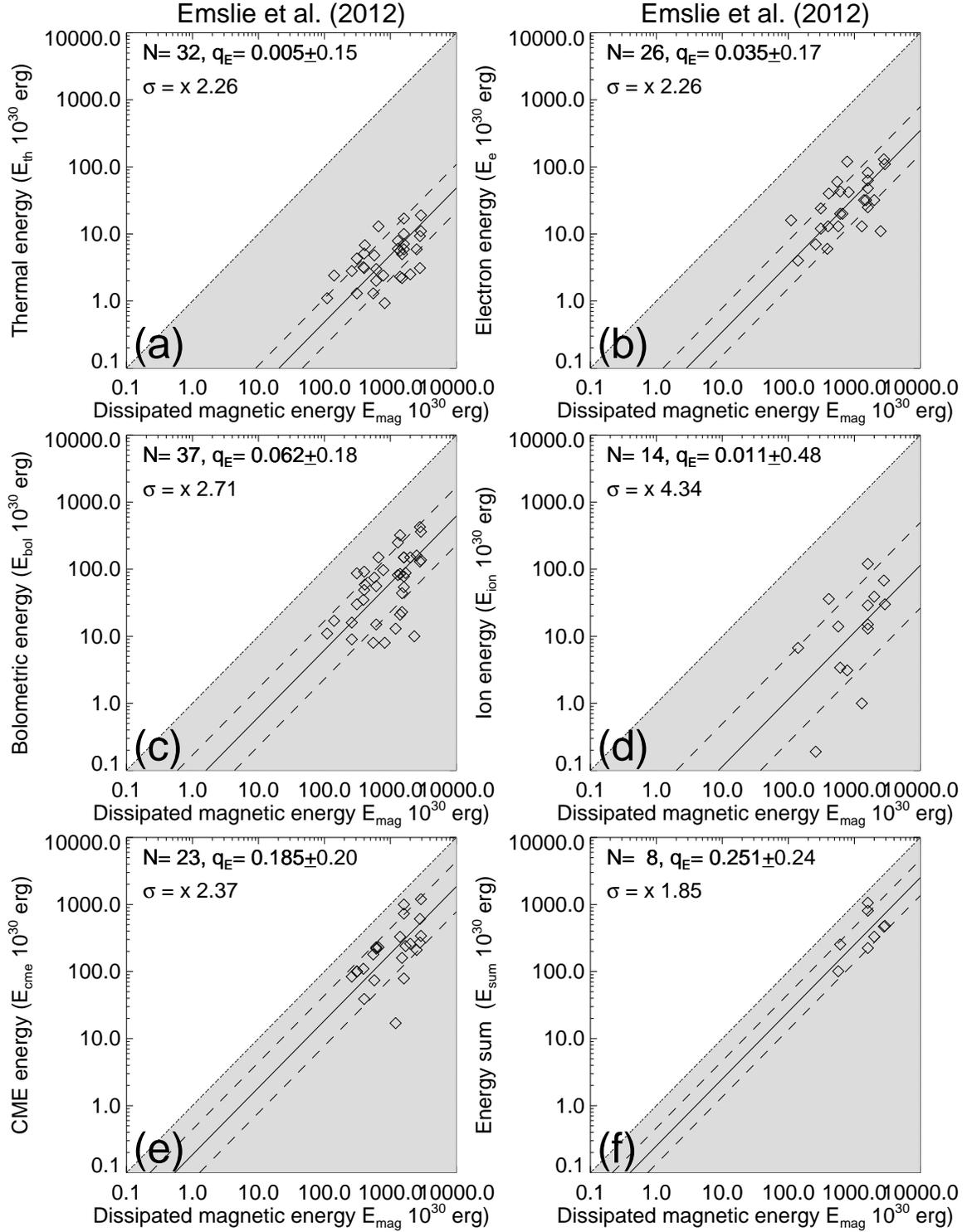}
\caption{Cross-correlation plots of the same parameters as shown
in Fig.~3, but for the data set of 37 eruptive flare events 
analyzed in Emslie et al.~(2012).}
\end{figure}

\begin{figure}
\plotone{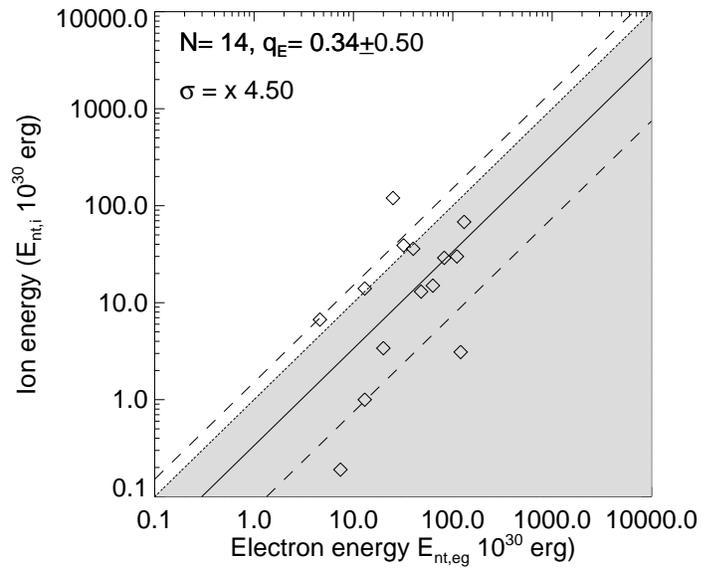}
\caption{Cross-correlation between nonthermal ion energies $E_{\mathrm{nt,i}}$
versus the nonthermal electron energies $E_{\mathrm{nt,e}}$ from a dataset
of 37 eruptive flare events analyzed in Emslie et al.~(2012).}
\end{figure}

\begin{figure}
\plotone{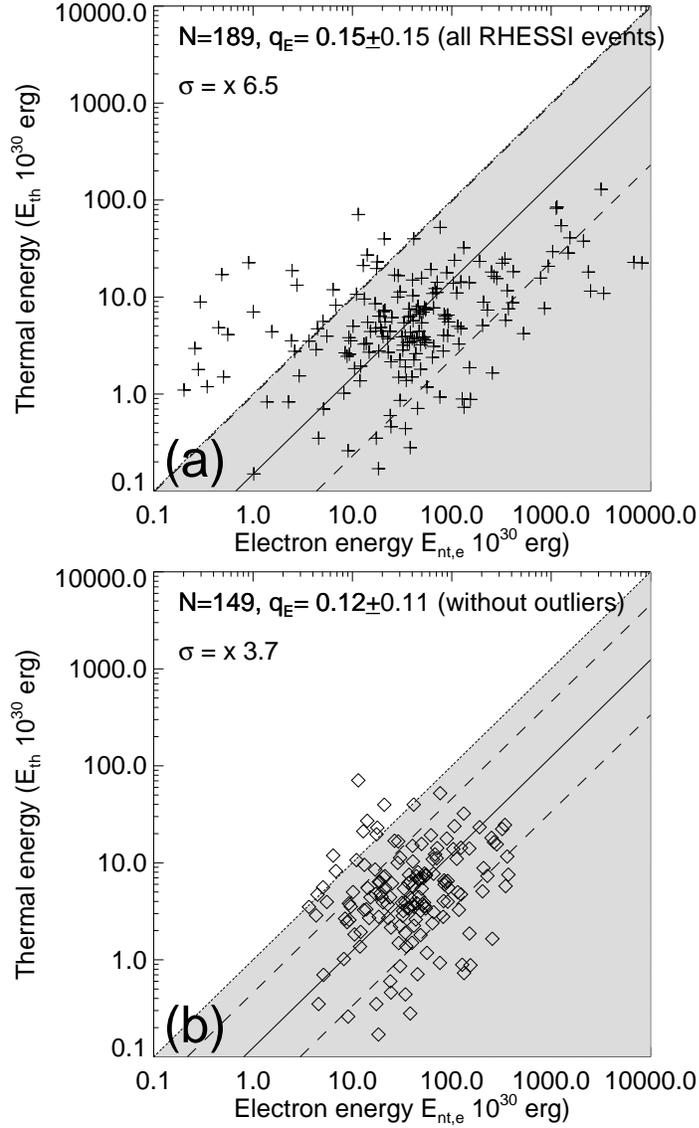}
\caption{Cross-correlation between thermal energies $E_{\mathrm{th}}$
and nonthermal electron energies $E_{\mathrm{nt,e}}$, for all {\it RHESSI}
events (top panel), and for a subset without outliers 
(bottom panel) according to Fig.~2b. Note that the thermal energy 
generally does not exceed the nonthermal energy (equivalence
is indicated with a diagonal bordering the grey zone).}
\end{figure}

\begin{figure}
\plotone{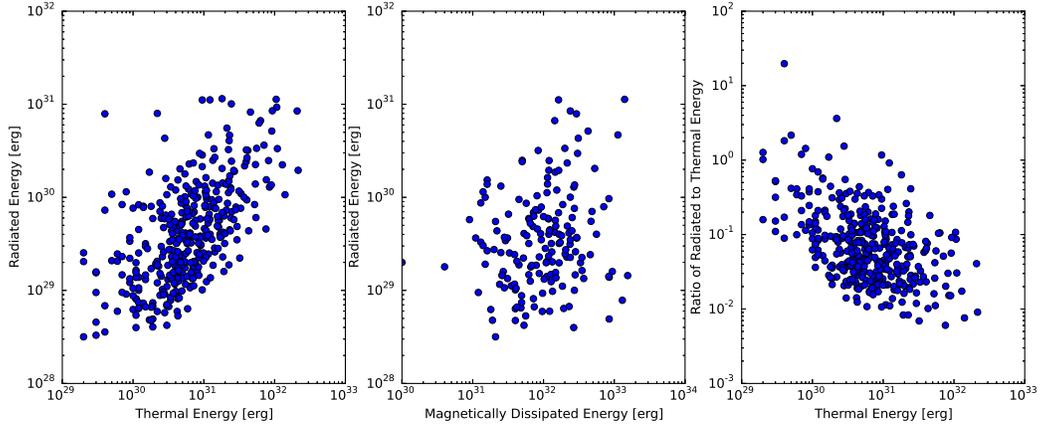}
\caption{Thermally radiated energies from the hot ($>$4 MK) 
coronal plasma as a function of thermal and magnetically dissipated 
energies: (a) Radiated energy versus thermal energy; (b) Radiated energy 
versus magnetically dissipated energies; and (c) Ratio of radiated to thermal 
energies versus thermal energy.}
\end{figure}

\begin{figure}
\plotone{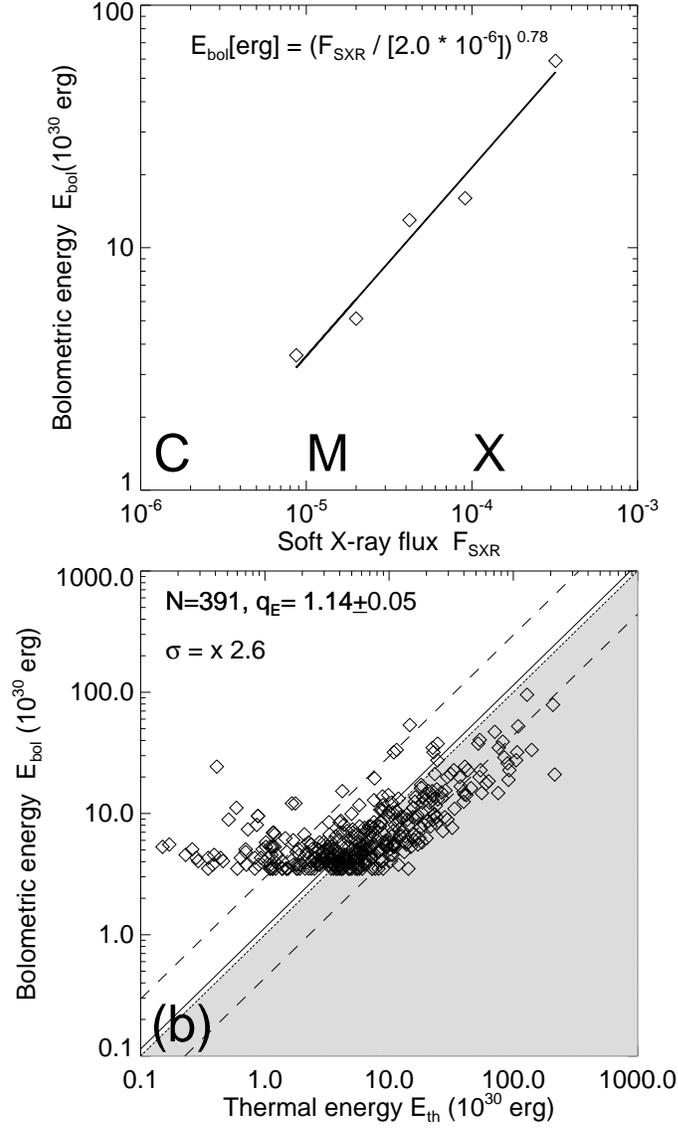}
\caption{Cross-correlation between bolometric energy $E_{\mathrm{bol}}$ and
the SXR flux $F_{\mathrm{SXR}}$ of the {\it GOES} 1-8 \ang\ flux according
to Kretzschmar (2011) (top panel). The resulting correlation between
the bolometric energy $E_{\mathrm{bol}}$ and the thermal energy 
$E_{\mathrm{th}}$ yields a mean ratio of almost unity.}
\end{figure}

\begin{figure}
\plotone{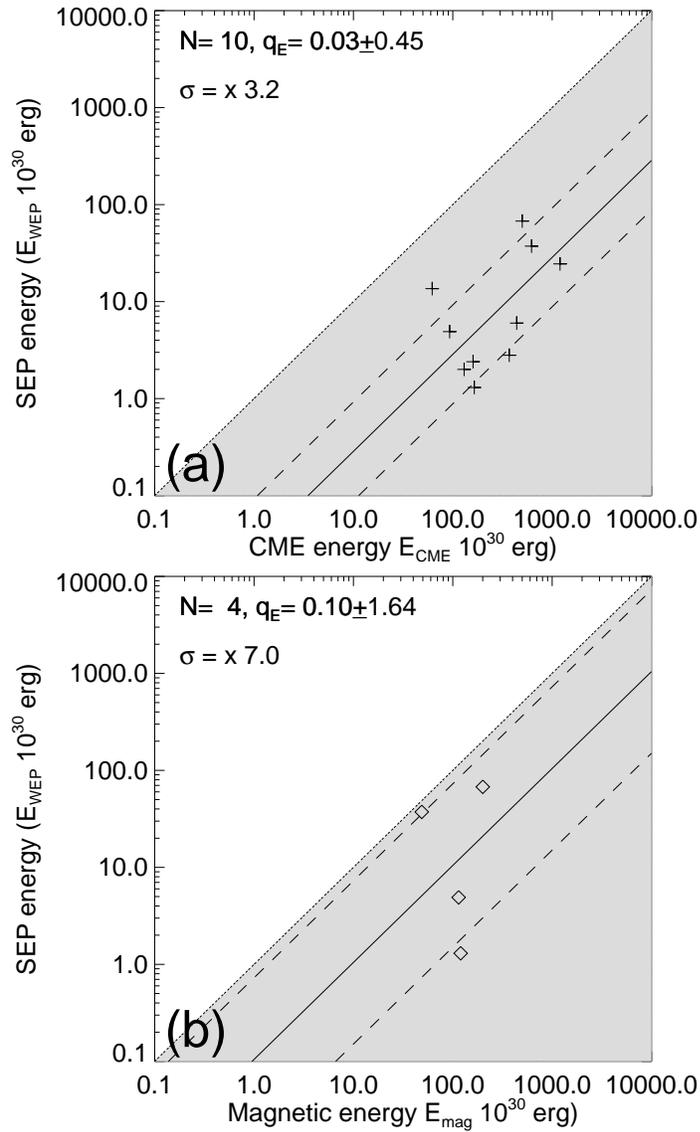}
\caption{The ratio of SEP kinetic energies $E_{\mathrm{SEP}}$ to CME energies
$E_{\mathrm{CME}}$ (a) and versus the dissipated magnetic energy
$E_{\mathrm{mag}}$ in flares (b), based on the SEP data
given in Table 1.}
\end{figure}

\begin{figure}
\plotone{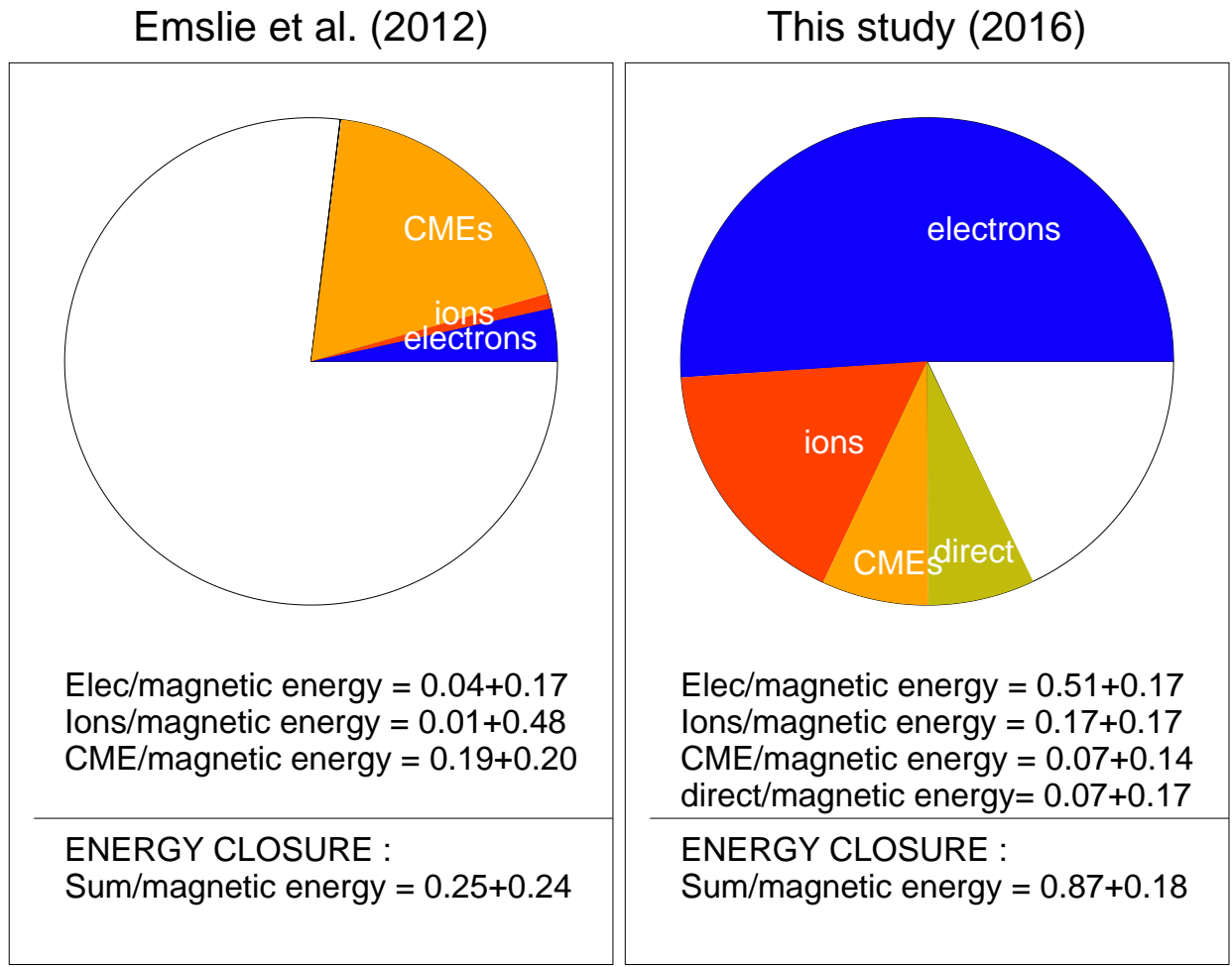}
\caption{Pie chart of energy closure, obtained from previous work of
Emslie et al.~(2012) (left panel), and from this study (right panel).}
\end{figure}

\begin{figure}
\plotone{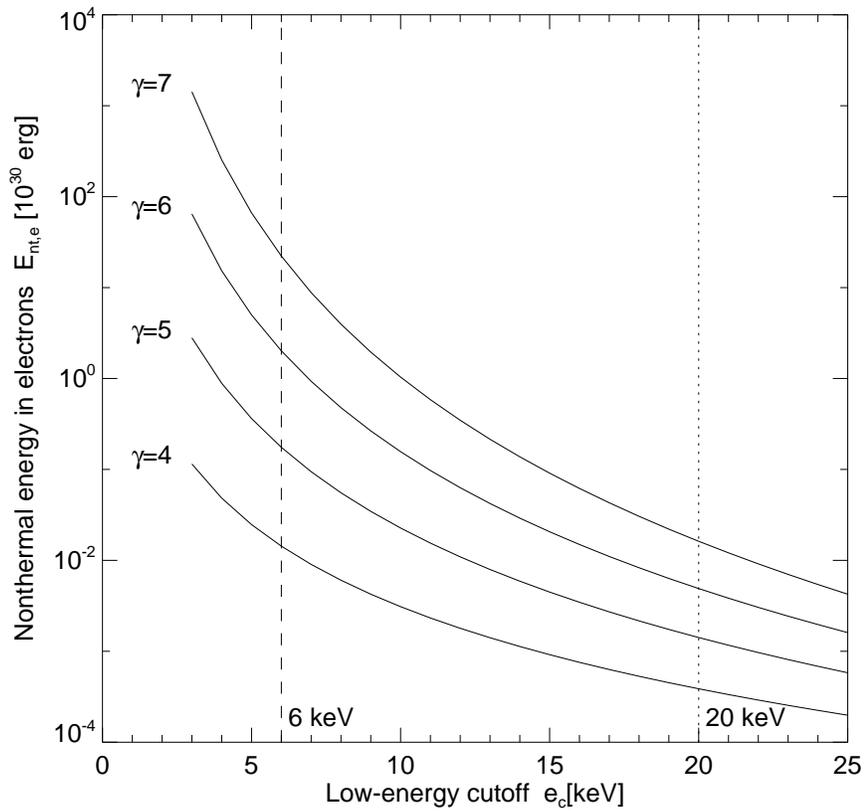}
\caption{The dependence of the nonthermal energy in electrons
$E_{\mathrm{nt,e}}$ on the low-energy cutoff $e_c$, calculated for four
different power law slopes $(\gamma=4-7)$ of the hard X-ray
photon spectrum. Two typical low-energy cutoffs are marked:
6 keV assumed for the warm-target model, and 20 keV as typical value
of the cross-over energy (Paper III).}
\end{figure}

\end{document}